\documentclass[twocolumn,pre,reprint,showpacs,preprintnumbers,amsmath,amssymb,superscriptaddress]{revtex4}
\usepackage{graphicx}% Include Fig.~files
\usepackage{dcolumn}% Align table columns on decimal point
\usepackage{bm}% bold mathf
\usepackage{amsmath}
\usepackage{physics}
\usepackage{multirow}
\newcommand{\tens}[1]{%
	\mathbin{\mathop{\otimes}\limits_{#1}}%
}

\let\oldAA\AA
\renewcommand{\AA}{\text{\normalfont\oldAA}}
\usepackage{upgreek}
\usepackage{eps2pdf}
\usepackage{amsmath}
\usepackage{subfigure}
\usepackage{hyperref}
\usepackage[normalem]{ulem}
\hypersetup{
    colorlinks,
    citecolor=blue,
    filecolor=black,
    linkcolor=blue,
    urlcolor=blue
}
\usepackage{epstopdf}
 \usepackage{relsize}

\newcommand*{\rom}[1]{\expandafter\@slowromancap\romannumeral #1@}
\title{Report}

\setcitestyle{square}

\begin{document}

%\preprint{APS/123-QED}

\title{Non-local triple quantum dot thermometer  based on Coulomb-coupled systems }% Force line breaks with \\
%\thanks{A footnote to the article title}%

%\email{sagnik.banerjee01@gmail.com}

% \altaffiliation[Also at ]{Physics Department, XYZ University.}%L3ines break automatically or can be forced with \\

	%This line break forced with \textbackslash\textbackslash
%
\author{Aniket Singha}

\thanks{email: aniket@ece.iitkgp.ac.in}

% \altaffiliation[Also at ]{Physics Department, XYZ University.}%L3ines break automatically or can be forced with \\

\affiliation{%
Department of Electronics and Electrical Communication Engineering, \\Indian Institute of Technology Kharagpur, Kharagpur-721302, India\\
 %This line break forced with \textbackslash\textbackslash
}%

%\email{aniket@ece.iitkgp.ac.in}
%\email{aniket@ece.iitkgp.ac.in}
%\affiliation{
%Department of Mechanical Engineering,\\
%Massachusetts Institute of Technology, 77, Massachusetts Avenue, Cambridge, MA 02139 \\
 %This line break forced with \textbackslash\textbackslash
%}%

%\author{Bhaskaran Muralidharan}%
%\email{bm@ee.iitb.ac.in}
%\affiliation{%
%Department of Electrical Engineering,\\
%Indian Institute of Technology Bombay, Powai, Mumbai-400076, India\\
 %This line break forced with \textbackslash\textbackslash
%}%

%\collaboration{MUSO Collaboration}%\noaffiliation

%\author{Charlie Author}
% \homepage{http://www.Second.institution.edu/~Charlie.Author}
%\affiliation{
 %Second institution and/or address\\
 %This line break forced% with \\
%}%
%\affiliation{
% Third institution, the second for Charlie Author
%}%
%\author{Delta Author}
%\affiliation{%
% Authors' institution and/or address\\
% This line break forced with \textbackslash\textbackslash
%}%

%\collaboration{CLEO Collaboration}%\noaffiliation
%\date{\today}% It is always \today, today,
             %  but any date may be explicitly specified

\begin{abstract}
 Recent proposals towards non-local thermoelectric voltage-based thermometry, in the conventional dual quantum dot set-up, demand an asymmetric step-like system-to-reservoir coupling around the  ground states for optimal operation (Physica E, 114, 113635, 2019).  In addition to such demand for unrealistic coupling, the sensitivity in such a strategy also depends on the average measurement terminal temperature, which may result in erroneous temperature assessment. In this paper, I propose non-local current based thermometry in the dual dot set-up as a practical alternative  and  demonstrate that in the regime of high bias, the  sensitivity  remains robust against fluctuations of the measurement terminal temperature. Proceeding further, I propose a non-local  triple quantum dot thermometer, that provides an enhanced sensitivity while bypassing the demand for unrealistic step-like  system-to-reservoir coupling   and being robust against fabrication induced variability in Coulomb coupling. In addition, I show that the heat  extracted from (to) the target reservoir, in the triple dot design, can also be suppressed drastically by appropriate fabrication strategy, to prevent thermometry induced drift in reservoir temperature. The proposed triple  dot setup  thus offers a multitude of benefits and could potentially pave the path towards the practical realization and deployment of high-performance non-local ``sub-Kelvin range" thermometers.

\end{abstract}
\maketitle
\section{Introduction}
Nanoscale electrical thermometry in the cryogenic domain, particularly in the sub-Kelvin regime,  has been one of the greatest engineering challenges in the current era. Device engineering with the ambition to couple system thermal parameters  with electrically measurable quantities has been extremely challenging in nano-scale regime. In the recent era of nano-scale engineering,  thermal manipulation of electron flow has manifested itself in the proposals of thermoelectric engines \cite{thermoelectric_transport_at_mesoscopic_level_1,aniket_nonlocalheat,thermoelectric_transport_at_mesoscopic_level_2,heatengine1,heatengine2,heatengine3,heatengine4,heatengine5,aniket,heatengine6,bd1,bd2,bd3,aniket_heat1,aniket_heat2,heatengine7,heatengine8,heatengine9,heatengine10,heatengine11,heatengine12,heatengine13,heatengine14,heatengine15,heatengine16,heatengine17}, refrigerators \cite{thermalrefrigerator1,thermalrefrigerator2,aniket_cool1,aniket_cool2,thermalrefrigerator3,thermalrefrigerator4,thermalrefrigerator5,thermalrefrigerator6,bd4,thermalrefrigerator7,thermalrefrigerator8}, rectifiers \cite{thermalrectifier1,thermalrectifier2,thermalrectifier3,thermalrectifier4,thermalrectifier5,thermalrectifier6} and transistors \cite{thermaltransistors1,thermaltransistors2,thermaltransistors3,thermaltransistors4,thermaltransistors5,thermaltransistors6,thermaltransistors7,thermaltransistors8}. In addition, the possibility of  \emph{non-local} thermal control of electrical parameters has been also been proposed and demonstrated experimentally \cite{3terminalheatengine1,3terminalheatengine2,3terminalheatengine3,3terminalheatengine4,3terminalheatengine5,3terminalheatengine6,3terminalheatengine7,3terminalheatengine8,3terminalheatengine9,3terminalrefrigerator1,3terminalrefrigerator2,3terminalrefrigerator3,3terminalrefrigerator4}. In the case of non-local thermal control, electrical parameters between two terminals are dictated by  temperature of one or more remote reservoirs,   which are spatially and electrically isolated from the path of current flow. The electrical and spatial isolation thus prohibits any exchange of electrons between the remote reservoir(s) and the current conduction track, while still permitting the reservoir(s) to act as the heat source (sink) via appropriate Coulomb coupling \cite{3terminalheatengine1,3terminalheatengine2,3terminalheatengine3,3terminalheatengine4,3terminalheatengine5,3terminalheatengine6,3terminalheatengine7,3terminalheatengine8,3terminalheatengine9,3terminalrefrigerator1,3terminalrefrigerator2,3terminalrefrigerator3,3terminalrefrigerator4}.    \\
\indent Thus,  non-local thermal manipulation of electronic flow mainly manifests itself in multi-terminal devices, where current/voltage between two terminals may be controlled via temperature-dependent stochastic fluctuation at one (multiple) remote electrically isolated  reservoir(s) \cite{3terminalheatengine1,3terminalheatengine2,3terminalheatengine3,3terminalheatengine4,3terminalheatengine5,3terminalheatengine6,3terminalheatengine7,3terminalheatengine8,3terminalheatengine9,3terminalrefrigerator1,3terminalrefrigerator2,3terminalrefrigerator3,3terminalrefrigerator4}. Non-local coupling between electrical and thermal parameters provides a number of distinct benefits over their local counterparts, which encompass isolation of the remote target reservoir from current flow induced  Joule heating, the provision of  independent engineering and manipulation  of electrical and lattice thermal conductance,  etc. 
\indent Recently proposals towards non-local thermometry via thermoelectric voltage measurement in a capacitively coupled dual quantum dot set-up  \cite{sensor_qdot} and current measurement in a point contact set-up \cite{sensor_qpc} have been put forward in literature. In such systems, the temperature of a remote target reservoir may be assessed via measurement of thermoelectric voltage or current  between two terminals that are electrically isolated from the target reservoir \cite{sensor_qdot,sensor_qpc}.  In addition, a lot of  effort has been directed towards theoretical and experimental demonstration of ``sub-Kelvin  range" thermometers \cite{thermometer1,thermometer2,thermometer3,thermometer4,milikelvinthermometer1,milikelvinthermometer2,milikelvinthermometer3,milikelvinthermometer4,milikelvinthermometer5,milikelvinthermometer6,milikelvinthermometer7,milikelvinthermometer8,milikelvinthermometer9,milikelvinthermometer10,milikelvinthermometer11}.  \\
\indent  In this paper, I first argue that non-local thermoelectric voltage based sensitivity in the conventional dual dot set-up, proposed in Ref.~\cite{sensor_qdot}, is dependent on the average temperature of the measurement terminals, which might affect temperature assessment. Following this, I illustrate that non-local current-based thermometry offers an alternative and robust approach where the sensitivity remains unaffected by the average temperature of the measurement terminals. Although current based thermometry in the dual dot set-up  \cite{sensor_qdot} offers an attractive alternative, the optimal performance of such a set-up demands a sharp step-like transition in the system-to-reservoir coupling, which is hardly achievable in reality. Hence, I  propose a triple quantum dot based non-local thermometer that can perform optimally, while circumventing the demand for any energy resolved change in the system-to-reservoir coupling. Although such a system is asymmetric and prone to non-local thermoelectric action, I show that its thermometry remains practically unaffected in the regime of high bias voltage. The performance and  operation regime of the triple dot thermometer is investigated and compared with the conventional dual dot set-up to demonstrate that the triple dot thermometer offers enhanced temperature sensitivity along with a reasonable efficiency,  while bypassing the demand for unrealistic step-like system-to-reservoir coupling and providing robustness against fabrication induced variability in Coulomb coupling. It is also demonstrated that the heat-extraction from the remote (non-local) target reservoir \cite{heatengine5,thermalrefrigerator5} in the triple dot set-up can be substantially suppressed,  without affecting the system sensitivity, by tuning  the dot to remote reservoir coupling. Thus the triple dot thermometer hosts a multitude of advantages, making it suitable for its realization and  deployment in practical applications.  \\
\indent This paper is organized as follows. In Sec. \ref{performance_parameters}, I discuss the parameters employed to gauge the performance of the thermometers. Next, Sec.~\ref{results} first illustrates  current-based non-local thermometry in the dual dot setup as an attractive alternative to thermoelectric voltage based operation  \cite{sensor_qdot}. Proceeding further in Sec.~\ref{results}, I illustrate and investigate the triple quantum dot based non-local thermometer.  Finally, I conclude the paper briefly in Sec. \ref{conclusion}. The derivation of the quantum master equations (QME) for the triple dot thermometer is given in the Appendix section.
 \begin{figure*}
 	\includegraphics[width=.9\textwidth]{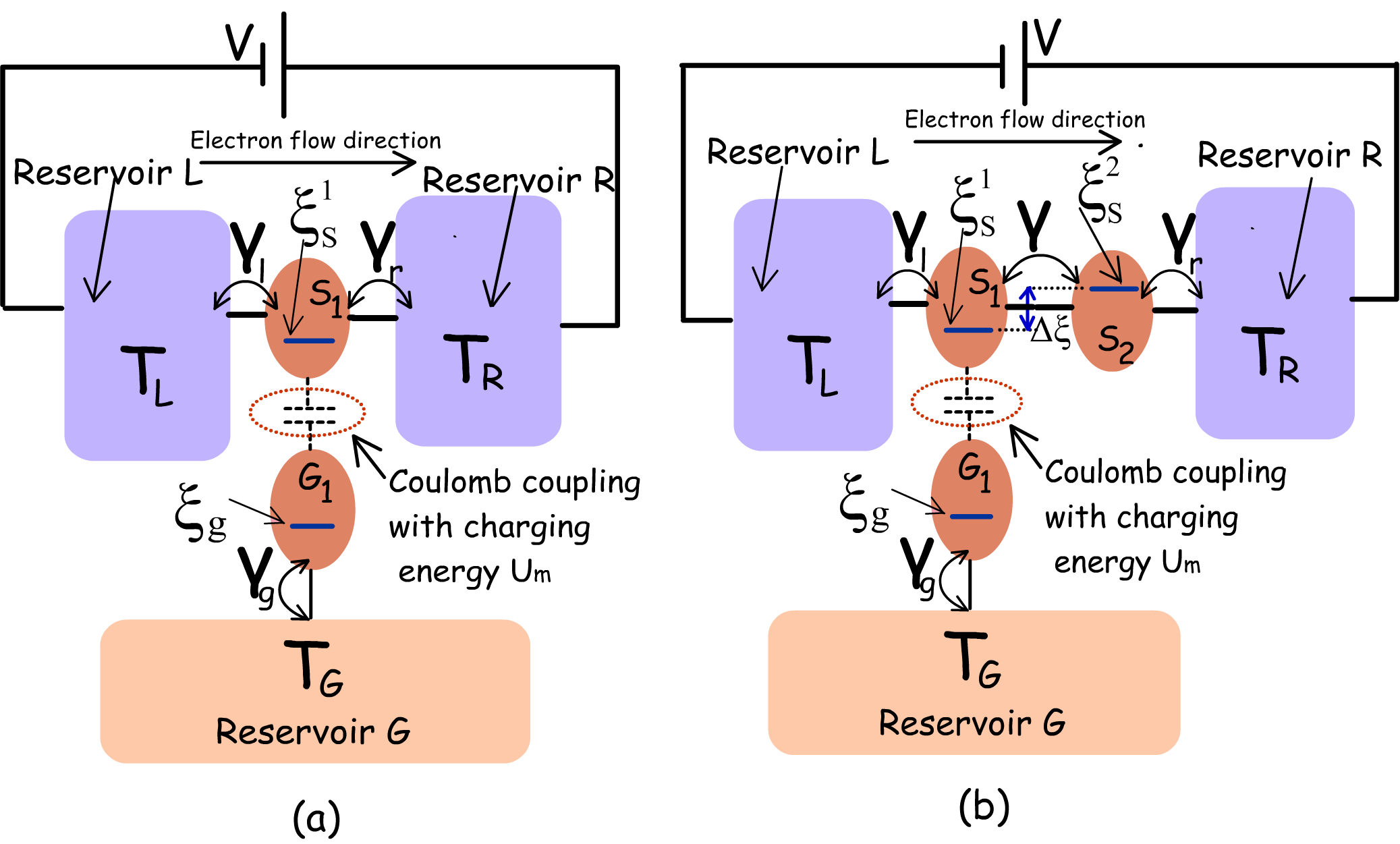}
 	\caption{Schematic of the dual dot and triple dot thermometer (a)  Schematic diagram of the dual dot thermometer based on Coulomb-coupled systems \cite{sensor_qdot}.   This thermometer set-up is based on a simpler thermodynamic engine proposed by S\'anchez \emph{et. al.} \cite{heatengine5} and consists of two Coulomb-coupled quantum dots $S_1$ and $G_1$. $S_1$ is electrically connected to the reservoirs $L$ and $R$ and provides the path for current flow. $G_1$ on the other hand, is electrically connected to the remote reservoir $G$ whose temperature is to be accessed. To investigate the optimal performance of the dual dot thermometer, I choose $\gamma_L(\xi)=\gamma_c \theta (\xi_s^1+\delta \xi-\xi)$, $\gamma_R(\xi)=\gamma_c\theta (\xi-\xi_s^1-\delta \xi)$ and $\gamma_g(\xi)=\gamma_c$ \cite{heatengine5} with $\gamma_c=10\upmu$eV. Here, $\theta$ is the Heaviside step function and $\delta \xi<U_m$.  (b) Schematic diagram of the proposed triple dot  electrical thermometer. The entire system consists of the dots $S_1$, $S_2$ and $G_1$, which  are electrically coupled to reservoirs \textit{L}, \textit{R}, and \textit{G} respectively.  $S_1$ and $G_1$ are capacitively coupled to each other (with Coulomb-coupling energy $U_m$). The ground state energy levels of the three dots $S_1$, $S_2$ and $G_1$ are  denoted by    $\xi_s^1$, $\xi_s^2$ and $\xi_g$ respectively. $S_1$ and $S_2$ share a staircase ground state configuration with $\xi_s^2=\xi_s^1+\Delta \xi$.  To assess  the optimal performance of the triple dot thermometer, I choose $\Delta E=U_m$ and   $\gamma_L(\xi)=\gamma_r(\xi)=\gamma_g(\xi)=\gamma_c$, with $\gamma_c=10\upmu$eV.}% This thermometer is a derivative of the non-local thermodynamic engine set-up proposed by S\'anchez, \emph{et. al.} \cite{thermaltransistors4}. }
 	\label{fig:Fig_1}
 \end{figure*}
 \section{Performance parameters of the thermometers}\label{performance_parameters}
 \indent The two types of non-local thermometers recently proposed in literature include (i) open-circuit voltage based thermometers \cite{sensor_qdot}, and (ii) current based thermometers \cite{sensor_qpc}. Both of these thermometers rely on Coulomb coupling. The parameter employed to gauge the thermometer performance should be related to the rate of change of an electrical variable with temperature and is termed as sensitivity. As such, sensitivity is defined as the rate of change in (i)   open-circuit voltage with temperature $\left(
 \frac{dV_o}{dT_G}\right)$ for voltage based thermometry and, (ii) current with temperature  $\left(
\chi= \frac{dI}{dT_G}\right)$ for current based thermometry. Here, $T_G$ is the remote target reservoir temperature  to be assessed. When it comes to current based thermometry, a second parameter of importance, related to the efficiency, may be defined as the sensitivity per unit power dissipation, which I term as the performance coefficient. Thus, performance coefficient is given by:
\begin{equation}
Performance-coefficient=\frac{\chi}{P},
\end{equation}
where $P=V\times I$ is the power dissipated across the set-up. In the above equation, $I$ indicates the current flowing through the thermometer on application of bias voltage $V$. It should be noted that the performance coefficient is a parameter to gauge the sensitivity with respect to power dissipation and is not a true efficiency parameter in sense of energy conversion. 
  \begin{figure}[!htb]
 	\includegraphics[width=0.5\textwidth]{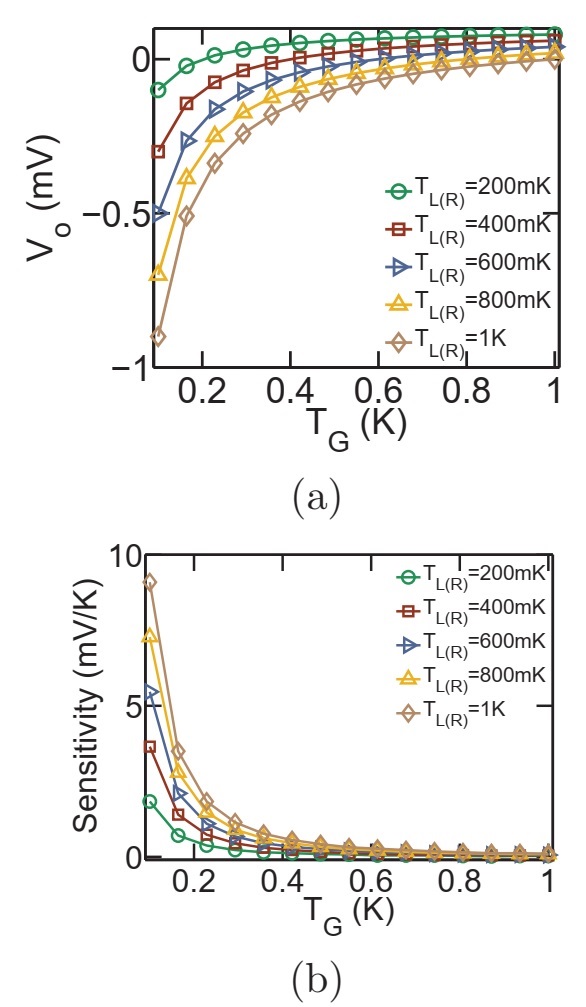}
 	\caption{Voltage based thermometry in the dual-dot set-up depicted in Fig.~\ref{fig:Fig_1}(a). Variation in (a) Open-circuit voltage (b) temperature sensitivity $\left(\frac{dV_o}{dT_G}\right)$ with $T_G$ for different values of $T_{L(R)}$. For the above set of plots, the value of Coulomb coupling energy is chosen as $U_m=100\upmu$eV and the ground states  are pinned at the equilibrium Fermi energy, that is, $\xi_s^1=\xi_g=\mu_0$. The open-circuit voltage as well as temperature  sensitivity $\left(\frac{dV_o}{dT_G}\right)$ in the set-up under consideration is dependent on $T_{L(R)}$. }
 	\label{fig:Fig_2}
 \end{figure} 
\begin{figure}[!htb]
	\includegraphics[width=0.45\textwidth]{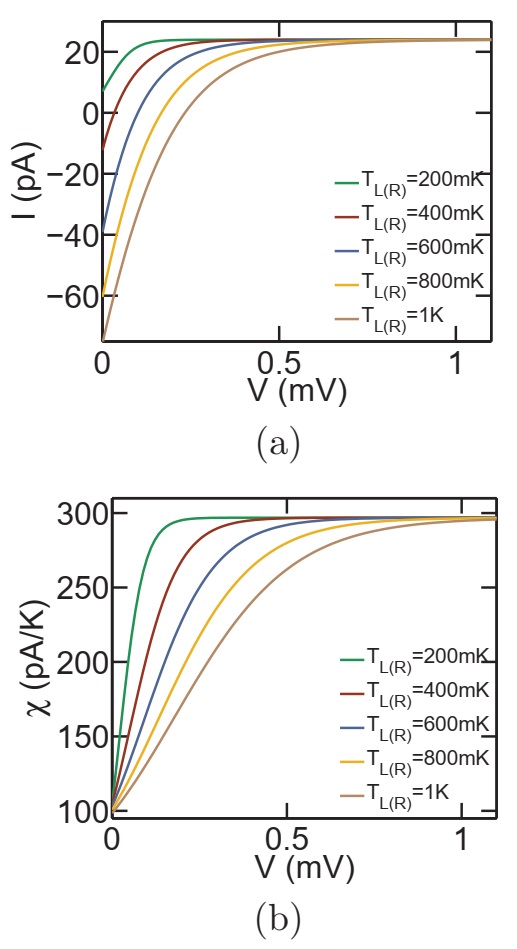}
	\caption{Current based thermometry in the dual-dot set-up depicted in Fig.~\ref{fig:Fig_1}(a). Variation in (a) total current (b) temperature sensitivity $\left(\chi=\frac{dI}{dT_G}\right)$ with applied bias $V$ for different values of $T_{L(R)}$. For the above set of plots, the parameters employed are $U_m=100\upmu$eV and $T_G=300$mK, while the ground states of $S_1$ and $G_1$ are pinned at the equilibrium Fermi energy, that is $\xi _s^1=\xi_g=\mu _0$. Given sufficiently high bias voltage $V$, the total current as well as  temperature sensitivity $\chi=\left(\frac{dI}{dT_G}\right)$  saturate to the same value for different $T_{L(R)}$. }
	\label{fig:Fig_3}
\end{figure}
\section{Results}\label{results}
\indent In this section, I investigate non-local open-circuit voltage and current based thermometry in the dual dot set-up. Proceeding further, I propose a triple dot design that demonstrates a superior sensitivity while circumventing the demand for any change in the system-to-reservoir coupling. In addition, the triple dot thermometer also demonstrates robustness against fabrication induced variability in Coulomb coupling.  The performance and operation regime in case of current based sensitivity for both the dual dot and the triple dot thermometers were investigated and compared.    The last part of this section investigates the thermometry induced refrigeration (heat-up) of the remote reservoir in the  dual and triple dot set-up and also elaborates a strategy to reduce such undesired effect in case of the triple dot design.   \\

\subsection{Thermometry in the dual dot set-up}\label{dd}
\indent The dual dot thermometer, schematically demonstrated in Fig.~\ref{fig:Fig_1}(a), is based on the non-local thermodynamic engine originally conceived by S\'anchez, \emph{et. al.} \cite{heatengine5}. It consists of two quantum dots $S_1$ and $G_1$. The dot $S_1$ is electrically tunnel coupled to reservoirs \textit{L} and  \textit{R}, while $G_1$ is electrically coupled to the reservoir \textit{G}. Here, $G$ is the target reservoir whose temperature is to be assessed. The temperature of the reservoirs $L,~R$ and $G$ are symbolized as $T_L,~T_R$ and $T_G$ respectively. The dots $S_1$ and $G_1$ are capacitively coupled with Coulomb coupling energy $U_m$, which permits exchange of electrostatic energy between the dots $S_1$ and $G_1$ while prohibiting any  flow of electrons between them, resulting in \textit{zero} net electronic current out of (into) the reservoir $G$. Thus the reservoir $G$ is electrically isolated from the current flow path. The ground state energy levels of the  dots $S_1$ and  $G_1$ are indicated by  $\xi_s^1$ and  $\xi_g$  respectively.  Due to mutual Coulomb coupling between $S_1$ and $G_1$, the change in electron number $n_{S_1}~(n_{G_1})$ of the dot $S_1$ ($G_1$)  influences the electrostatic energy of the  dot $G_1$ ($S_1$). Under the assumption that the change in potential   due to self-capacitance  is much greater than the applied voltage $V$ or the average thermal voltage $kT/q$, that is  $q^2/C_{self}>>(qV, k_BT)$, the electron occupation probability or transfer rate via the Coulomb
blocked energy level, due to self-capacitance can be neglected \cite{heatengine5}. Thus, the system analysis can be approximated by considering four multi-electron levels by limiting the maximum number of electrons in the ground state of each quantum dot to $1$. Denoting each state by the electron occupation number in the quantum dot ground state, a possible system state of interest may be represented as $\ket{n_{S_1},n_{G_1}}=\ket{n_{S_1}}\tens{} \ket{n_{G_1}}$, where $n_{S_1},n_{G_1}\in (0,1)$,  denote the number of electrons present in the ground-states of $S_1$ and $G_1$ respectively. The above  assumptions validate the use of the  quantum master equations (QME) employed  in Ref.~\cite{heatengine5} for investigation of an equivalent set-up. It was demonstrated in Refs.~\cite{heatengine5,sensor_qdot}  that optimal operation of the dual-dot based set-up as heat engine and thermometer demands an asymmetric step-like system-to-reservoir coupling. Hence, to investigate the optimal performance of the dual dot thermometer, I choose $\gamma_l(\xi)=\gamma_c \theta (\xi_s^1+\delta \xi-\xi)$and $\gamma_r(\xi)=\gamma_c\theta (\xi-\xi_s^1-\delta \xi)$ \cite{heatengine5} with $\gamma_c=10\upmu$eV and  $\delta \xi<U_m$. Here, $\theta$ and $\xi$ respectively are the Heaviside step function and the free-variable denoting energy. In addition, I choose $\gamma_g=\gamma_c$. Such order of coupling parameter  correspond to realistic experimental values in Ref. \cite{heatengine13}, where the system-to-reservoir coupling was evaluated, from experimental data, to lie in the range of $20\sim 50\upmu$eV.  In addition, such order of the coupling parameters also indicate  weak  coupling and limit the electronic transport in the sequential tunneling regime where the impact of cotunneling and higher-order tunneling processes can be neglected. Unless stated, the temperature of the reservoirs $L$ and $R$ are assumed to be $T_{L(R)}=300$mK. To assess the performance of the thermometer, I follow the approach as well as the quantum master equations employed in Refs. \cite{heatengine5,thermalrefrigerator5}, where the probability of occupancy of the considered multi-electron states were evaluated via well established quantum master equations (QME) to finally calculate the charge and heat currents through the system. \\
\indent   \textit{\textbf{Voltage-based thermometry:}} In case of non-local thermoelectric voltage based thermometry, the applied bias $V$ in Fig.~\ref{fig:Fig_1}(a) is replaced by open circuit and the voltage between the terminals $L$ and $R$ is measured. Such  open circuit voltage based thermometry  for the considered dual dot set-up  was analyzed earlier in detail by Zhang, \emph{et. al.} \cite{sensor_qdot}. I plot, in Fig.~\ref{fig:Fig_2}, the variation in open-circuit voltage ($V_o$) and temperature sensitivity $\left(\frac{dV_o}{dT_G}\right)$ for different values of $T_{L(R)}$ at $U_m=100\upmu$eV. It is evident that the open-circuit voltage as well as  sensitivity $\left(\frac{dV_o}{dT_G}\right)$ in such a set-up is dependent on $T_{L(R)}$, which makes it non-robust against fluctuations in the measurement terminal temperature. The variation in  open-circuit voltage and sensitivity with $T_{L(R)}$ results from the fact that non-local thermoelectric voltage developed in such set-ups is dependent on $\Delta T =T_{L(R)}-T_G$.\\
 \begin{figure*}[!htb]
\includegraphics[width=1\textwidth]{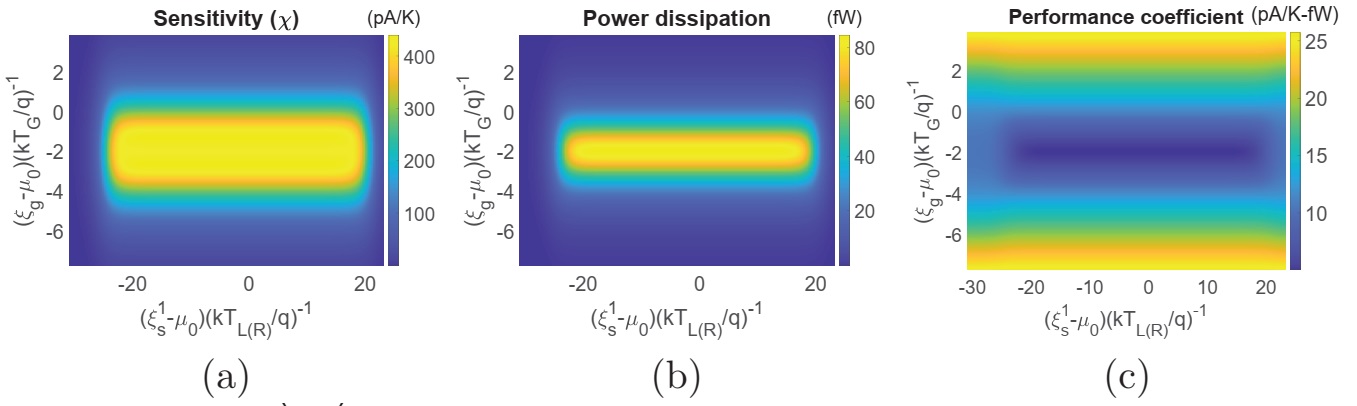}
	\caption{Regime of operation of the dual dot set-up in terms of the ground state energy positions relative to the equilibrium Fermi energy $\mu_0$. Colour plot demonstrating the variation in (a) sensitivity ($\chi$) (b) power dissipation and (c) performance coefficient with variation in the ground state positions $\xi_s^1$ and $\xi_g$. The parameters used for simulation are $U_m=100\upmu$eV, $\gamma_c=10\upmu$eV, $V=1.1$mV and  $T_{L(R)}=T_G=300$mK.}
	\label{fig:Fig_4}
\end{figure*}
\indent  \textit{\textbf{Current-based thermometry:}} To ensure robustness in such a set-up against fluctuation and variation in measurement terminal temperature and voltage, current based thermometry offers an alternative  method. In this case, a bias voltage $V$ is applied between the reservoirs $L$ and $R$ and temperature of the reservoir $G$ can be assessed via  the current measurement. As stated before, temperature sensitivity in this case is defined as 
\begin{equation}
\chi=\frac{dI}{dT_G},
\end{equation}
where $I$ is the electronic current flowing between the reservoirs $L$ and $R$. Fig.~\ref{fig:Fig_3} demonstrates the variation in electronic current $I$ and temperature sensitivity $\chi=\left(\frac{dI}{dT_G}\right)$ for different values of $T_{L(R)}$ at $U_m=100\upmu$eV.  It should be noted that the set-up is affected by non-local thermoelectric action in the regime of low bias, which is evident from different magnitudes of current at distinct values of $T_{L(R)}$. However, for sufficiently high bias voltage, the electronic current as well as the sensitivity $\chi=\left(\frac{dI}{dT_G}\right)$ saturate to a finite limit for different values of $T_{L(R)}$. Thus, in the regime of high bias, current based thermometry in the set-up under consideration is robust against thermoelectric effect, fluctuations in the bias voltage and variation in measurement terminal temperature $T_{L(R)}$. \\
\indent Fig.~\ref{fig:Fig_4} demonstrates the regime of operation of the set-up under consideration with respect to the ground state energy positions for $U_m=100\upmu$eV, $V=1.1$mV and $T_{L(R)}=T_G=300$mK. Such values of the applied bias drive the thermometer in the regime of maximum saturation sensitivity. In particular, Fig.~\ref{fig:Fig_4}(a) demonstrates the variation in sensitivity ($\chi$) with  ground state positions $\xi_s^1$ and $\xi_g$ relative to the equilibrium Fermi level $\mu_0$. We note that the optimal sensitivity is obtained when $\xi_g$ lies within the range of a few $kT_G$ below the equilibrium Fermi energy $\mu_0$. This is because, the flow of an electron from reservoir $L$ to $R$ demands the entry of an electron in dot $G_1$ at energy $\xi_g+U_m$ and subsequently exit of the electron from $G_1$ into reservoir $G$  at an energy $\xi_g$ \cite{heatengine5,thermalrefrigerator5}. To understand this, let us consider the complete cycle that transfers an electron from reservoir $L$ to $R$ in the dual dot set-up: $\ket{0,0} \rightarrow \ket{1,0} \rightarrow \ket{1,1} \rightarrow \ket{0,1} \rightarrow \ket{0,0}$. In this cycle, an electron tunnels into the dot $S_1$ from reservoir $L$ at energy $\xi_s^1$. Next, an electron tunnels into the dot $G_1$ from reservoir $G$ at energy $\xi_g+U_m$. In the following step, the electron in $S_1$ tunnels out into the reservoir $R$ at energy $\xi_s^1+U_m$. The system returns to the vacuum state, that is $\ket{0,0}$ when the electron in $G_1$ tunnels out into reservoir $G$ at energy $\xi_g$.  Thus, the  sensitivity becomes optimal in the regime around  the maximum value of the factor $\frac{d}{dT_G}\left[f\left(\frac{\xi_g+U_m-\mu_0}{kT_G}\right)\left\{1-f\left(\frac{\xi_g-\mu_0}{kT_G}\right)\right\}\right]$, which occurs when $\xi_g $ is a few $kT_G$ below the equilibrium Fermi energy $\mu_0$. Similarly, the power dissipation, shown in Fig.~\ref{fig:Fig_4}(b),  is high when $\xi_g$ lies within the range of a few $kT_G$ below the equilibrium Fermi energy $\mu_0$ due to high current flow. Interestingly, by comparing Fig.~\ref{fig:Fig_4}(a) and (b), we  find regimes where the sensitivity is high at a relatively lower power dissipation.   The performance coefficient (shown in Fig.~\ref{fig:Fig_4}.c), on the other hand, is low in the regime of high sensitivity and increases as $\xi_g$ deviates from the equilibrium Fermi energy beyond a few $kT_G$. This can be explained as follows. In the regime of high sensitivity, the current flow is high. Due to limited current carrying capacity of the dual dot set-up, the rate of fractional increase in current flow  with $T_G$, that is $\left(\frac{1}{I}\frac{dI}{dT_G}\right)$, is lower in the regime of high current flow. Hence, although the sensitivity is high, the rate of fractional increase in current flow with temperature, and hence the sensitivity per unit power dissipation is lower. This gives rise to low performance coefficient. On the other hand, in the regime of low sensitivity, the current flow is lower (evident from the lower power dissipation). Thus, the rate of fractional increase in current flow  with $T_G$, that is $\left(\frac{1}{I}\frac{dI}{dT_G}\right)$, is higher in this regime. This gives rise to high performance coefficient  in the regime of low sensitivity. From Fig.~\ref{fig:Fig_4}(a-c), we also note that the sensitivity, power dissipation and performance coefficient is fairly constant over a wide range of $\xi_s^1$. Although not shown here, this range depends on  and increases (decreases) with the increase (decrease) in the applied bias voltage. \\
 \begin{figure*}[!htb]
 	\includegraphics[width=1\textwidth]{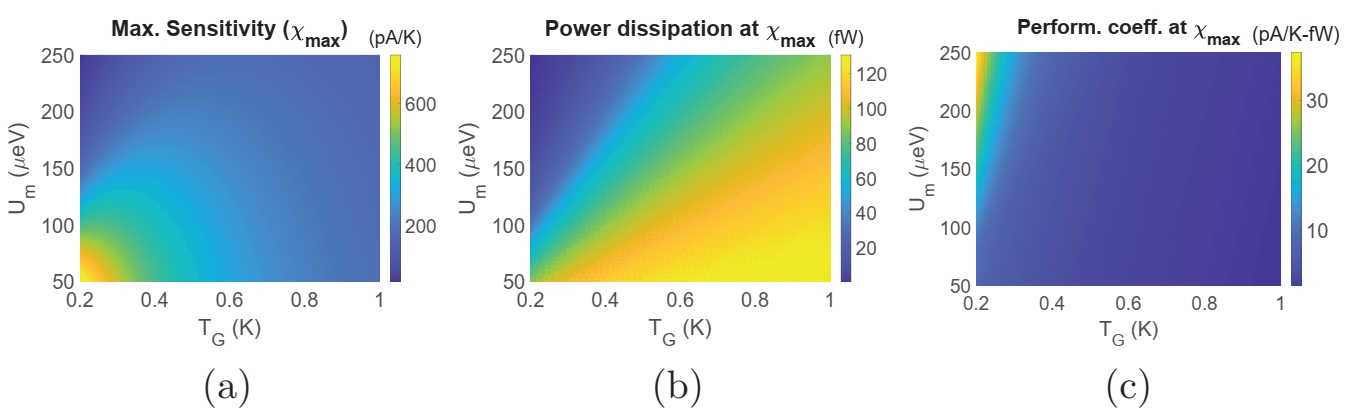}
 	\caption{Maximum sensitivity and parameters at maximum sensitivity for the dual dot thermometer. Colour plot demonstrating the variation in (a) maximum sensitivity ($\chi_{max}$) (b) power dissipation at maximum sensitivity and (c) performance coefficient at maximum sensitivity with variation in the Coulomb coupling energy ($U_m$) and target reservoir temperature ($T_G$). The parameters used for simulation are $T_{L(R)}=300$mK, $\gamma_c=10\upmu$eV and  $V=1.1$mV.}
 	\label{fig:Fig_5}
 \end{figure*}
\indent I demonstrate in Fig.~\ref{fig:Fig_5}, the variation in maximum sensitivity ($\chi_{max}$), as well as, power dissipation and performance coefficient at the maximum sensitivity with variation in the Coulomb coupling energy ($U_m$) and $T_G$ respectively. To calculate the maximum sensitivity and related parameters at the maximum sensitivity, the ground states are tuned to optimal positions with respect to the equilibrium Fermi energy ($\mu_0$). We note that the maximum sensitivity, shown in Fig.~\ref{fig:Fig_5}(a), is relatively higher in the regime of low Coulomb coupling energy $U_m$ and decreases with  $U_m$. This is because the maximum value of $\frac{d}{dT_G}\left[f\left(\frac{\xi_g+U_m-\mu_0}{kT_G}\right)\left\{1-f\left(\frac{\xi_g-\mu_0}{kT_G}\right)\right\}\right]$ decreases with increase in $U_m$. Moreover, we also note that the sensitivity changes non-monotonically with $T_G$. Coming to the aspect of power dissipation, we note that the dissipated power at the maximum sensitivity decreases monotonically with $U_m$. This, again,  is due to decrease in the optimal value of the  product $f\left(\frac{\xi_g+U_m-\mu_0}{kT_G}\right)\left\{1-f\left(\frac{\xi_g-\mu_0}{kT_G}\right)\right\}$ with $U_m$, which results in decrease in the current flow and, hence power dissipation. In addition, the power dissipation also increases with  $T_G$ for the same reason of increase in current due to increase in the product $f\left(\frac{\xi_g+U_m-\mu_0}{kT_G}\right)\left\{1-f\left(\frac{\xi_g-\mu_0}{kT_G}\right)\right\}$ with $T_G$. The performance coefficient at the maximum sensitivity, as noted from Fig.~\ref{fig:Fig_5}(c), is maximum in the regime of low temperature and high Coulomb coupling energy $U_m$, rendering this set-up suitable for applications in the ``sub-Kelvin" temperature regime.

\subsection{Thermometry in triple-dot set-up}
 \begin{figure*}[!htb]
	\includegraphics[width=1\textwidth]{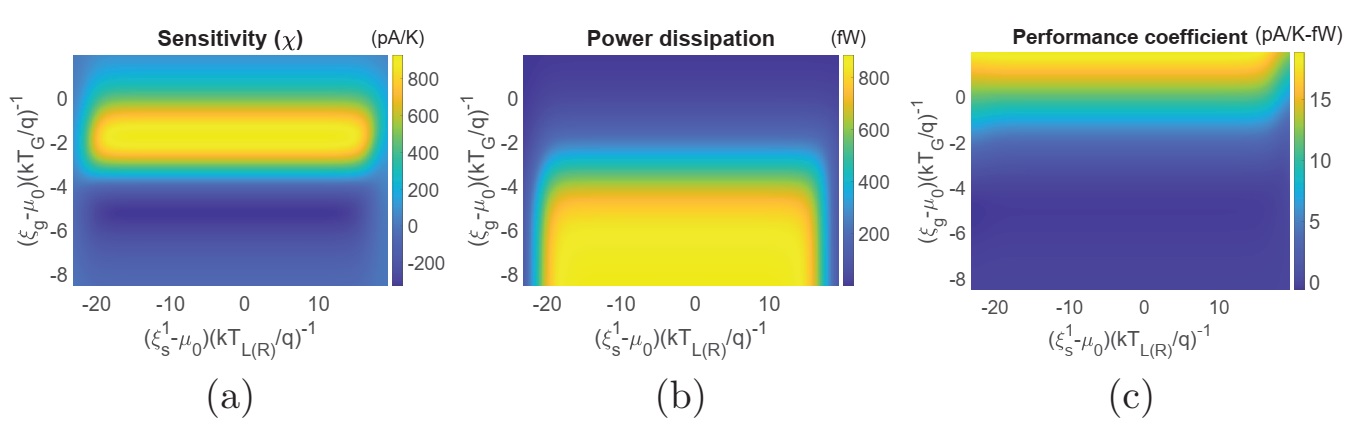}
	\caption{Regime of operation of the proposed triple dot thermometer in terms of the ground state energy positions relative to the equilibrium Fermi energy $\mu_0$. Colour plot demonstrating the variation in (a) sensitivity ($\chi$) (b) power dissipation and (c) performance coefficient with variation in the ground state positions $\xi_s^1$ and $\xi_g$. The parameters used for simulation are $U_m=100\upmu$eV, $\gamma_c=10\upmu$eV, $V=1.1$mV and  $T_{L(R)}=T_G=300$mK.}
	\label{fig:Fig_6}
\end{figure*}
\indent   \textit{\textbf{Proposed set-up configuration and transport formulation:}} The triple dot thermometer, proposed in this paper, is schematically demonstrated in Fig.~\ref{fig:Fig_1}(b) and consists of three dots $S_1,~S_2$ and $G_1$ which  are electrically coupled to the reservoirs $L$, $R$ and $G$ respectively. Compared to the dual-dot design, the triple dot set-up   features an extra quantum dot $S_2$ between $S_1$ and reservoir $R$.  Coming to the ground state configuration and other features of the system, $S_1$ and $S_2$ are tunnel coupled to each other, while $G_1$ is capacitively coupled to $S_1$. The ground states of $S_1$ and $S_2$ form a stair-case configuration with $\xi_s^2= \xi_s^1+\Delta \xi$.   Any electronic tunneling between the dots  $S_1$ and $G_1$ is suppressed via suitable fabrication techniques \cite{cap_coup_1,cap_coup_2,cap_coup3,cap_coup_4,cap_coup_5}. Energy exchange between $S_1$ and $G_1$ is,  however, feasible via Coulomb coupling \cite{cap_coup_1,cap_coup_2,cap_coup3,cap_coup_4,cap_coup_5}. In the optimal dual-dot thermometer discussed above, an asymmetric step-like system-to-reservoir coupling is required for optimal operation. In the proposed triple-dot thermometer, the asymmetric system-to-reservoir coupling is bypassed by  choosing an energy difference between the ground states of  $S_1$ and $S_2$ which makes the system  asymmetric with respect to the reservoir $L$ and $R$.   Another equivalent triple-dot set-up, based on Coulomb coupled systems, that can be employed for efficient non-local thermometry is demonstrated in Fig.~\ref{fig:app_2} and discussed briefly in Appendix \ref{app_a}.   Coming to the realistic fabrication possibility of such a system, due to the recent advancement in solid-state nano-fabrication technology, triple and quadruple  quantum dot systems with and without Coulomb coupling have already been realized experimentally \cite{mqd1,mqd2,mqd3,mqd4,mqd5,mqd6}. In addition, it has been experimentally demonstrated that quantum dots that are far from each other in space, may be bridged to obtain strong Coulomb coupling, along with  excellent thermal isolation between the reservoirs which may be at different temperatures \cite{cap_coup_1,cap_coup_2,cap_coup3,cap_coup_4,cap_coup_5}. Also, the bridge may be fabricated between two specific quantum dots to drastically enhance their mutual Coulomb coupling, without affecting the electrostatic energy of the other quantum dots in the system \cite{cap_coup_1,cap_coup_2,cap_coup3,cap_coup_4,cap_coup_5}.  Due to mutual Coulomb coupling between $S_1$ and $G_1$, the change in electron number $n_{S_1}~(n_{G_1})$ of the dot $S_1$ ($G_1$)  influences the electrostatic energy of the  dot $G_1$ ($S_1$).
The total increase in electrostatic energy $U$ of the triple dot configuration,  demonstrated in Fig.~\ref{fig:Fig_1} (b),  due to deviation in electronic number from the minimum energy configuration can be given by (Appendix \ref{app_b}): 
\begin{widetext}
	\begin{equation}
	U(n_{S_{1}},n_{G_{1}},n_{S_{2}})=\sum_{x }U^{self}_{x}\left(n_{x}^{tot}-n_x^{eq}\right)^2   +\sum_{(x_{1},x_{2})}^{x_1 \neq x_2} U^m_{x_1,x_2}\left(n_{x_1}^{tot}-n_{x_1}^{eq}\right)\left(n_{x_2}^{tot}-n_{x_2}^{eq}\right) 
	\end{equation}
\end{widetext}
where $n_x^{tot}$  is the total  electron number, and $U^{self}_x=\frac{q^2}{C^{self}_{x}}$ is the electrostatic energy due to self-capacitance $C^{self}_{x}$ (with the surrounding leads) of    quantum dot `$x$' (See Appendix \ref{app_b} for details). $U^m_{x_1,x_2}$ is the electrostatic energy arising out of  Coulomb coupling between two different quantum dots $x_1$ and $x_2$, where a change in electron number in $x_1$ affects the electrostatic energy in $x_2$ or vice-versa (Appendix \ref{app_b}). $n_x^{eq}$ is the total number of electrons in dot $x$ under equilibrium condition at $0K$ and is determined by the lowest possible electrostatic energy of the system. Hence, $n_x=n_x^{tot}-n_x^{eq}$ is the total number of  electrons  in the ground state of the dot $x$ due to application of bias voltage or stochastic fluctuations from the reservoirs  (Details given in Appendix \ref{app_b}).  Under the assumption that the change in potential  due to self-capacitance is much greater than than the average thermal voltage $kT/q$ or the applied bias voltage $V$, that is $U^{self}_x=\frac{q^2}{C^{self}_{x}}>> (kT,~qV)$, the electron occupation probability or transfer rate via the Coulomb blocked energy level, due to self-capacitance, is negligibly small.  Under such a condition, the analysis  of the triple dot system  can be approximated by limiting the maximum number of electrons in each  dot to unity. So, there are $2^3=8$ multi-electron  levels which characterize the entire non-equilibrium properties of the set-up. I denote each of these states by the ground state occupation number in each quantum dot. Hence, a possible state of interest in the system may be denoted  as $\ket{n_{S_1},n_{G_1},n_{S_2}}=\ket{n_{S_1}}\tens{} \ket{n_{G_1}} \tens{} \ket{n_{S_2}}$, where $n_{S_1},n_{G_1},n_{S_2}\in (0,1)$,   denote the number of electrons present in the ground-states of $S_1,~G_1$ and $S_2$ respectively. I also  neglect electrostatic coupling between $S_1,~S_2$ and  $S_2,~G_1$    for all practical purposes under consideration. Due to mutual coupling,  the ground states as well as electronic transport  in $S_1$ and $G_1$ are inter-dependent and hence, I  treat the pair of dots $S_1$ and $G_1$ as a  sub-system ($\varsigma_1$), $S_2$ being the complementary sub-system ($\varsigma_2$) of the entire triple-dot set-up (Appendix~\ref{app_b}).  The state probability of   $\varsigma_1$ is denoted by $P_{i,j}^{\varsigma_1}$,  $i$ and $j$ being the ground state electron number of dot $S_1$ and $G_1$ respectively. $P_k^{\varsigma_2}$, on the other hand, denotes the probability of occupancy of the dot $S_2$ in the complementary sub-system $\varsigma_2$. It can be shown that if $\Delta \xi$ is much greater than the system-to-reservoir coupling, that is $\Delta \xi>>\gamma_c$, then the interdot tunneling rate between $S_1$ and $S_2$ becomes maximum under the condition $\xi_s^1+U^m_{S_1,G_1}=\xi_s^2$, that is when $\Delta \xi=U^m_{S_1,G_1}$  (Appendix \ref{app_b}). To investigate the optimal performance of the proposed thermometer, I hence assume $\Delta \xi=U^m_{S_1,G_1}$. In the following discussion, I would simply represent $U^m_{S_1,G_1}$ as $U_m$.  Under the  above set of assumptions, the  equations dictating sub-system steady-state probabilities are given by (Appendix \ref{app_b}):	
\small
\begin{widetext}
	\begin{align}
	&  -P_{0,0}^{\varsigma_1}\{f_L(\xi_s^1)+f_G(\xi_g)\}+P_{0,1}^{\varsigma_1}\{1-f_G(\xi_g)\}+P_{1,0}^{\varsigma_1}\{1-f_L(\xi_s^1)\}=0 \nonumber \\
	& -P_{1,0}^{\varsigma_1}\left\{1-f_L(\xi_{s}^1)+f_G(\xi_g+U_m)\right\}+P_{1,1}^{\varsigma_1}\left\{1-f_G(\xi_g+U_m)\right\}+P_{0,0}^{\varsigma_1}f_L(\xi_s^1) \nonumber \\
	&-P_{0,1}^{\varsigma_1}\left\{1-f_g(\xi_{g}^1)+f_L(\xi_s^1+U_m)+\frac{\gamma}{\gamma_c}P^{\varsigma_2}_1\right\}+P_{0,0}^{\varsigma_1}f_G(\xi_g)+P_{1,1}^{\varsigma_1}\left\{1-f_L(\xi_s^1+U_m)+\frac{\gamma}{\gamma_c}P^{\varsigma_2}_{0}\right\} = 0 \nonumber \\
	& -P_{1,1}^{\varsigma_1}\left\{[1-f_g(\xi_{g}^1+U_m)]+[1-f_L(\xi_s^1+U_m)]+\frac{\gamma}{\gamma_C}P^{\varsigma_2}_0\right\}+P_{1,0}^{\varsigma_1}f_G(\xi_g+U_m) +P_{0,1}^{\varsigma_1}\left\{f_L(\xi_s^1+U_m)+\frac{\gamma}{\gamma_c}P^{\varsigma_2}_{1}\right\}=0
	\label{eq:first_sys}
	\end{align} 
\end{widetext}
\begin{align}
& -P_{0}^{\varsigma_2}\{f_R(\xi_s^2)+\frac{\gamma}{\gamma_c}P_{1,1}^{\varsigma_1}\}+P_1^{\varsigma_2}\left\{1-f_R(\xi_{s}^2)+\frac{\gamma}{\gamma_c}P^{\varsigma_1}_{0,1}\right\}=0\nonumber \\
&-P_1^{\varsigma_2}\{1-f_R(\xi_{s}^2)+\frac{\gamma}{\gamma_c}P^{\varsigma_1}_{0,1}\}+P_{0}^{\varsigma_2}\left\{f_R(\xi_s^2)+\frac{\gamma}{\gamma_c}P_{1,1}^{\varsigma_1}\right\}=0,
\label{eq:second_sys}
\end{align}
\normalsize where $\gamma_l(\xi)=\gamma_r(\xi)=\gamma_g(\xi)=\gamma_c$ and $\gamma$ are related to the reservoir-to-system tunnel coupling and the inter-dot tunnel coupling respectively \cite{dattabook}, $\xi$ being the independent energy variable.  In the above set of equations, $f_{\lambda}(\xi)$ denotes the probability of occupancy of the reservoir $\lambda$ at energy $\xi$. For the purpose of  calculations in this paper, I assume a quasi-equilibrium Fermi-Dirac statistics at the reservoirs. Hence, $f_{\lambda}(\xi)$ is given by:
\begin{equation}
f_{\lambda}(\xi)=\left(1+exp\left\{\frac{\xi-\mu_{\lambda}}{kT_{\lambda}}\right\}\right)^{-1},
\end{equation} 
where $\mu_{\lambda}$ and $T_{\lambda}$ respectively denote the quasi-Fermi energy and temperature of the reservoir $\lambda$. From the set of Eqns.~\eqref{eq:first_sys} and \eqref{eq:second_sys}, it is clear that an electron in $S_1$ can tunnel into $S_2$ only when the ground state in the dot $G_1$ is occupied with an electron. The set of  Eqns.~\eqref{eq:first_sys} and \eqref{eq:second_sys} are coupled to each other and may be solved using any iterative method. Here, I use Newton-Raphson iterative method to solve the steady-state values of sub-system probabilities. On calculation of the sub-system state probabilities $P_{i,j}^{\varsigma_1}$ and $P_k^{\varsigma_2}$, the electron current flow into (out of) the system from the reservoirs $L ( R)$ can be given as:
\small
\begin{align}
I_L= & q\gamma_c \times \left\{P^{\varsigma_1}_{0,0}f_L(\xi_s^1)+P^{\varsigma_1}_{0,1}f_L(\xi_s^1+U_m)\right\} \nonumber \\ &- q\gamma_c P^{\varsigma_1}_{1,0}\{1-f_L(\xi_s^1)\}- q\gamma_c P^{\varsigma_s^1}_{1,1}\{1-f_L(\xi_s^1+U_m)\} \nonumber \\
I_R= & -q\gamma_c \times \left\{P^{\varsigma_2}_{0}f_R(\xi_s^1)-P^{\varsigma_2}_{1}\{1-f_R(\xi_s^1)\}\right\} ,
\label{eq:final}
\end{align}
\normalsize
In addition, the electronic component of heat flow from the  reservoir $G$ can be given by:
\small
\begin{equation}
I_{Qe}=U_m \gamma_c\left\{P^{\varsigma_1}_{10}f_G(\xi_g+U_m)-P^{\varsigma_1}_{11}\{1-f_G(\xi_g+U_m)\}\right\}
\label{eq:heat}
\end{equation}  
\normalsize
Interestingly, we note that Eqn.~\eqref{eq:heat} is not directly dependent on  $\xi_g$. This is due to the fact that the net electronic current into or out of the reservoir $G$ is zero (See Appendix \ref{app_b} for details). \\
 \begin{figure*}[!hbt]
	\includegraphics[width=1\textwidth]{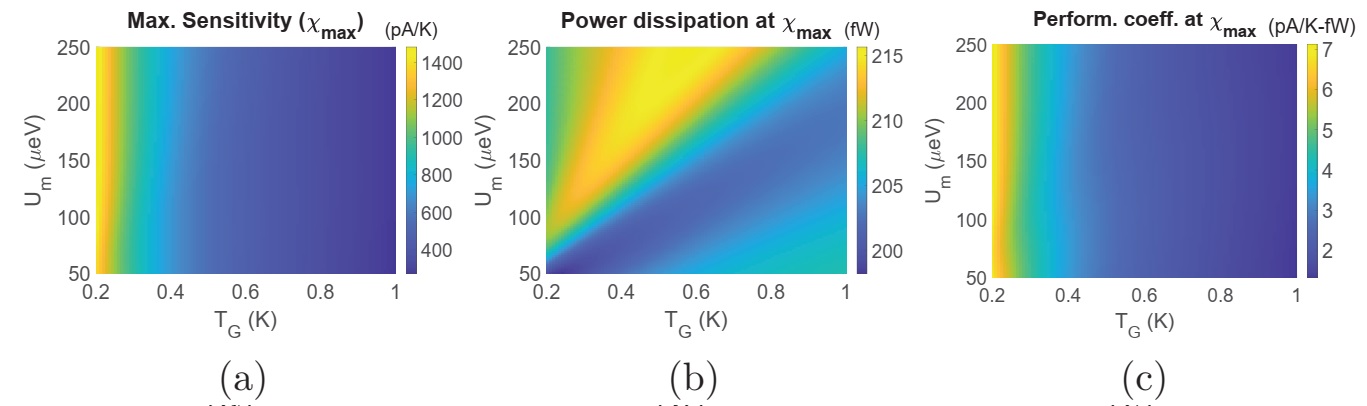}
	\caption{Maximum sensitivity and parameters at maximum sensitivity for the triple dot thermometer. Colour plot demonstrating the variation in (a) maximum sensitivity ($\chi_{max}$) (b) power dissipation at maximum sensitivity and (c) performance coefficient at maximum sensitivity with variation in the Coulomb coupling energy ($U_m$) and target reservoir temperature ($T_G$). The parameters used for simulation are $T_{L(R)}=300$mK,  $\gamma_l(\xi)=\gamma_r(\xi)=\gamma_g(\xi)=\gamma(\xi)=\gamma_c=10\upmu$eV and $V=1.1$mV.}
	\label{fig:Fig_7}
\end{figure*}
\indent \textbf{\textit{Operation regime and performance investigation:}} For investigating the triple dot set-up, I choose the system-to-reservoir coupling as $\gamma _l(\xi)=\gamma _r(\xi)=\gamma_g (\xi)=\gamma_c$, with  $\gamma_c=10\upmu$eV. In addition, I also choose the interdot coupling to be $\gamma(\xi)=10\upmu$eV. As stated earlier, such values of coupling parameters lie within experimentally feasible range \cite{heatengine13}. Fig.~\ref{fig:Fig_6} demonstrates the regime of operation of the proposed triple dot thermometer. In particular, Fig.~\ref{fig:Fig_6}(a) depicts the sensitivity as a function of the ground state positions. We note that the sensitivity increases as $\xi_g$ gradually goes below the Fermi energy, with the maximum sensitivity occurring when $\xi_g-\mu_0 \approx -1.5kT_G$. As $\xi_g$ goes further below the Fermi energy, the sensitivity becomes negative. This occurs when an increase in temperature decreases the probability of occupancy of both the ground state $\xi_g$ and the Coulomb blocked state $\xi_g+U_m$, that is when $\xi_g+U_m-\mu_0<0$. Despite the fact that this set-up offers the provision to implement a positively sensitive as well as a negatively sensitive thermometer, it should be noted from Fig.~\ref{fig:Fig_6}(b) that the power dissipation is very high in the negatively sensitive regime. This is due to the fact that when $\xi_g+U_m-\mu_0<0$, the occupancy probability of  $G_1$ is very high, which causes a high drive current between reservoirs $L$ and $R$. The power dissipation in the regime of positive sensitivity is lower, resulting in a higher performance coefficient, as noted from Fig.~\ref{fig:Fig_6}(c). Also, the power dissipation and performance coefficient respectively decreases and increases as $\xi_s^1$ gradually approaches and finally moves above the equilibrium Fermi-energy. This is because as $\xi_g$ gradually approaches and goes above the Fermi energy, the probability of occupancy of $\xi_g$ becomes lower, blocking the current flow through the system. Due to the same reason as stated for the dual dot set-up, a lower current flow through the system leads to a higher fractional increase in current  with the remote reservoir temperature $T_G$,  leading to a  higher performance coefficient. We also note from Fig.~\ref{fig:Fig_6}(a)-(c) that the sensitivity, power dissipation and performance coefficient remains almost constant for a wide range of $\xi_s^1$. As discussed before, this range depends on and increases (decreases) with increase (decrease) in applied bias voltage.\\
\begin{figure*}[!htb]
	\centering \includegraphics[width=.9\textwidth]{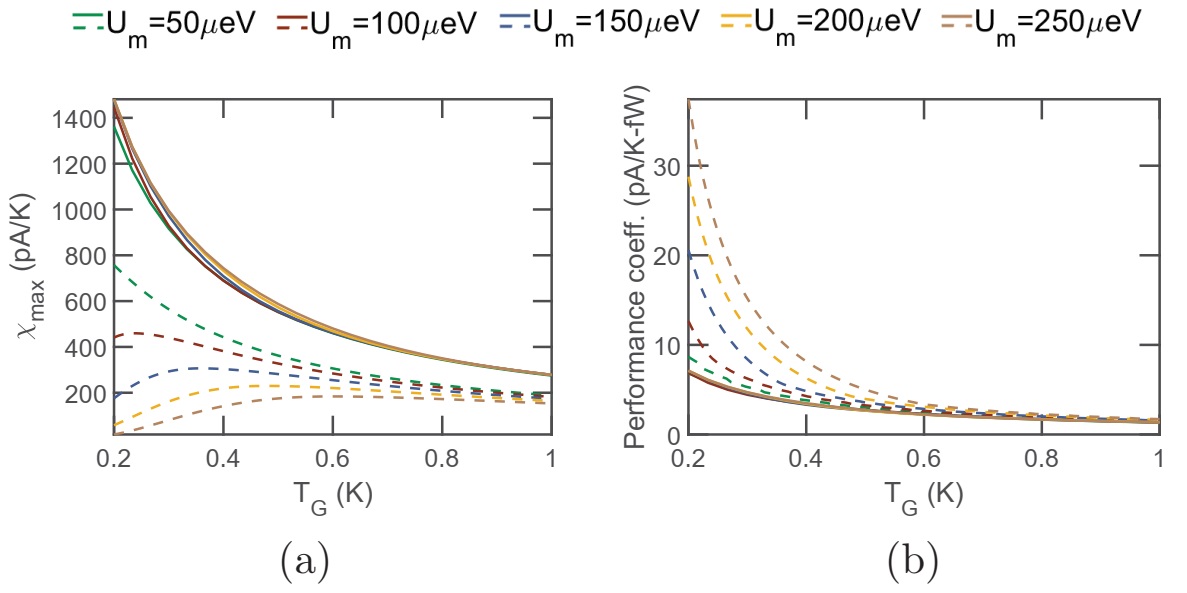}
	\caption{Performance comparison between the dual dot and the triple dot thermometer. Variation in (a) maximum sensitivity ($\chi_{max}$) and (b) Performance-coefficient at the maximum sensitivity with $T_G$ for different values of Coulomb coupling energy $U_m$. The solid and the dashed line represent the performance parameters of the triple dot and dual dot thermometers respectively.  The system parameters used for simulation are $T_{L(R)}=300$mK,   and $V=1.1$mV. For the dual dot thermometer, the different system to reservoir coupling are chosen to be $\gamma_l(\xi)=\gamma_c \theta (\xi_s^1+\delta \xi-\xi)$, $\gamma_r(\xi)=\gamma_c\theta (\xi-\xi_s^1-\delta \xi)$ and $\gamma_g=\gamma_c$. For the triple dot thermometer, the system to reservoir, as well as the interdot coupling are chosen to be $\gamma_l(\xi)=\gamma_r(\xi)=\gamma_g(\xi)=\gamma(\xi)=\gamma_c=10\upmu$eV. In both the dual dot and the triple dot thermometer, $gamma_c=10\upmu$eV.}
	\label{fig:Fig_8}
\end{figure*}
\begin{figure*}[!htb]
	\centering
	\centering \includegraphics[width=.9\textwidth]{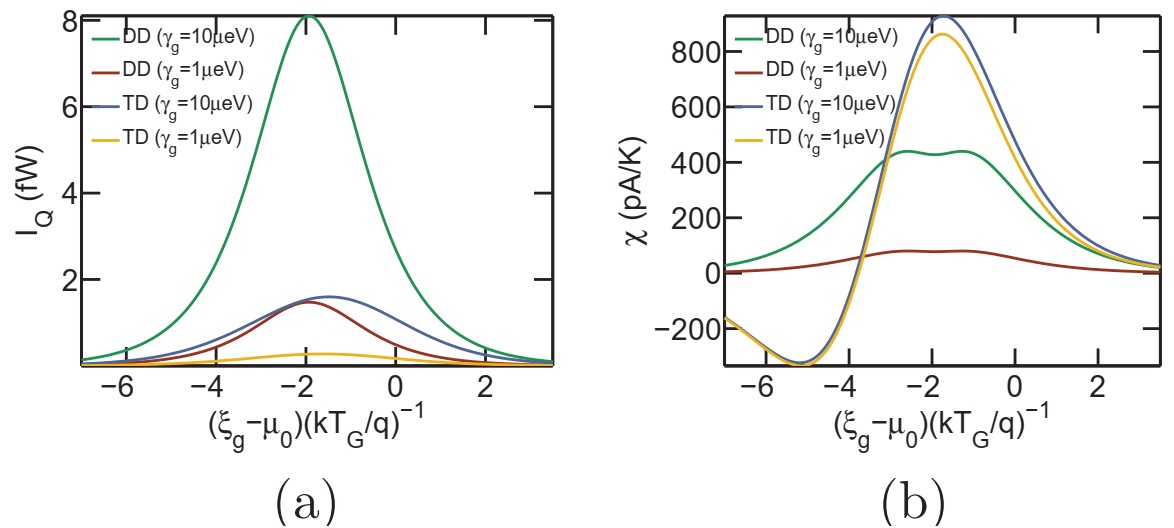}
	\caption{Analysis of thermometry induced refrigeration of the reservoir $G$ for the dual-dot (DD) and triple dot (TD) set-up.  Plot of (a) heat current ($I_Q$) extracted from the reservoir $G$ and (b) sensitivity with variation in the ground state $\xi_g$. In case of the dual dot (DD) set-up, decreasing the system-to-reservoir coupling ($\gamma_g$) between $G$ and $G_1$ decreases both the extracted heat current $I_Q$ and sensitivity $\chi$. However, for the triple dot set-up, decreasing $\gamma_g$ suppresses only the heat current $I_Q$, while keeping the sensitivity ($\chi$) almost unaltered. The parameters used for simulation are $U_m=100\upmu$eV, $\gamma_c=10\upmu$eV, $T_{L(R)}=T_G=300$mK and $\xi_s^1=\mu_0$.}
	\label{fig:Fig_9}
\end{figure*}
\indent Fig.~\ref{fig:Fig_7} demonstrates the maximum sensitivity ($\chi_{max}$) as well as the power dissipation and performance coefficient at the maximum sensitivity with variation in the Coulomb coupling energy $U_m$ and target reservoir temperature $T_G$. Just as before, to calculate the maximum sensitivity and related parameters at the maximum sensitivity, the quantum dot ground states are tuned to their optimal positions.  Fig.~\ref{fig:Fig_7}(a) demonstrates the maximum sensitivity with variation in $U_m$ and $T_G$. An interesting thing to note is that the triple dot thermometer is fairly robust against variation in the Coulomb coupling energy $U_m$. This can be explained by the fact that current flow through the triple quantum dot set-up only demands the occupancy of the dot $G_1$ whose ground state can be tuned to optimum position for maximizing the sensitivity. Thus, optimal sensitivity can be achieved by placing $\xi_g$ around the energy $\xi$ at which the rate of change in ground state occupancy probability of $G_1$ is maximum with $T_G$. This condition is unlike the case of dual dot set-up where one has to maximize the factor $\frac{d}{dT_G}\left[f\left(\frac{\xi_g+U_m-\mu_0}{kT_G}\right)\left\{1-f\left(\frac{\xi_g-\mu_0}{kT_G}\right)\right\}\right]$ for achieving the maximum sensitivity. We also note that, unlike the dual dot set-up, the maximum sensitivity in this case decreases monotonically with $T_G$. The power dissipation, as demonstrated in Fig.~\ref{fig:Fig_7}(b), also remains almost constant and varies between $199$fW and $216$fW with variation in $U_m$ and $T_G$. This again is a result of the fact that current flow through the triple dot set-up only demands occupancy of the dot $G_1$ and thus the position of $\xi_g$ for maximum sensitivity induces a high current flow through the set-up. Due to almost constant power dissipation with variation in $U_m$ and $T_G$, the performance-coefficient also shows a similar trend as the sensitivity with $U_m$ and $T_G$, as noted in Fig.~\ref{fig:Fig_7}(c). It is evident from Fig.~(\ref{fig:Fig_4})-(\ref{fig:Fig_7}) that the triple dot thermometer demonstrates an enhanced sensitivity, but lower performance coefficient compared to the dual dot thermometer. As such, it is important to compare their performance, which leads us to the next discussion. 
\subsection{Performance comparison}
To further shed light on the relative performance of the triple dot thermometer with respect to the dual dot thermometer, I plot in Fig.~\ref{fig:Fig_8}(a) and (b) the sensitivity and performance-coefficient respectively for the dual dot (dashed lines) and the triple dot (solid lines) thermometers respectively. As stated earlier, the triple dot thermometer demonstrates an enhanced sensitivity and  offers significant advantage, particularly in the regime of high Coulomb coupling energy $U_m$. This is due to the fact that each electronic flow between reservoirs $L$ and $R$ in the dual dot set-up demands an electron entrance and exit from $G_1$ at energy $\xi_g+U_m$ and $\xi_g$ respectively. Thus, the probability of electronic flow is significantly reduced, particularly for high $U_m$. Electronic flow in the triple dot set-up on the other hand demands only occupancy of the dot $G_1$, which can be achieved by positioning the ground state $\xi_g$ appropriately with respect to the equilibrium Fermi energy. Thus, this system eliminates the dependence of sensitivity on $U_m$, making it fairly robust against fabrication induced variability in the Coulomb coupling energy. The performance coefficient of the triple dot set-up, on the other hand, is lower compared to the dual dot thermometer. This is due to high current flow in the triple dot thermometer and becomes particularly noticeable in the regime of high values of $U_m$, where the dual dot set-up hosts very less current flow and sensitivity but high performance coefficient. It should be noted that the performance coefficient offered by the triple dot thermometer is reasonable and approaches  that of the dual dot set-up in the  higher temperature regime.
\subsection{Thermometry induced refrigeration}
 It is well known that the transfer of each electron from reservoir $R$ to $L$, in the dual dot set-up, demands extraction of a heat packet $U_m$ from reservoir $G$ \cite{heatengine5,thermalrefrigerator5}. This means that increasing the system-to-reservoir coupling to achieve enhanced sensitivity would also result in extraction of more heat packets from reservoir $G$. Such a phenomena may result in unnecessary  refrigeration or temperature drift of the reservoir $G$ in an undesirable manner. Since, the number of heat packets extracted in this set-up is exactly equal to the number of electrons that flow between reservoir $L$ and $R$ ($I_Q=IU_m/q$), reducing $\gamma_g$ to suppress the refrigeration of reservoir $G$ also results in the reduction of sensitivity. This is shown in Fig.~\ref{fig:Fig_9}(a) and (b), where it is demonstrated that reduction in $\gamma_g$ for the dual dot (DD) set-up, by a factor of $10$, results in suppression of both the maximum heat current  ($I_Q$) from $8.1$fW to $1.47$fW and maximum sensitivity ($\chi$) from $440$pA/K to $80$pA/K. Thus, both the maximum heat current and maximum sensitivity decrease by a factor of approximately $5.5$   \\
 \begin{figure}[!htb]
	\centering
	\centering \includegraphics[width=.5\textwidth]{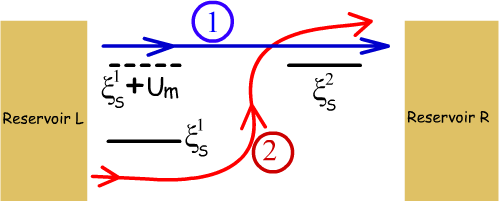}
	\caption{Schematic diagram depicting the two current components through the triple dot set-up. ``Component 1" (directed blue arrow) flows without absorbing heat packets from the remote reservoir $G$ and only depends on the occupancy probability of the ground state of $G_1$. ``Component 2"  (directed red line) flows by absorbing heat packets from the reservoir $G$, and results in extraction of heat  from the same.}
	\label{fig:Fig_E}
\end{figure}
\indent  In this aspect of refrigeration of the target reservoir $G$, the proposed triple dot set-up, on the other hand, offers a significant edge over the dual dot set-up. It should be noted that an electron flow in the triple dot set-up does not always demand the extraction of a heat packet from the reservoir $G$. To understand this, the components of current flow in the triple dot set-up are demonstrated in Fig.~\ref{fig:Fig_E}. As noted from Fig.~\ref{fig:Fig_E}, ``Component 1" flows directly from reservoir $L$ to $R$, without absorbing heat packets from reservoir $G$. This component flows when the ground state of the dot $G_1$ is occupied. Hence, it depends mainly on the probability of occupancy of the dot $G_1$ and is not directly controlled by the parameter $\gamma_g$. ``Component 2", on the other hand, flows when the electron enters in the dot $S_1$ with unoccupied  ground state of the dot $G_1$. Hence, this component flows by absorbing heat packets from reservoir $G$ and depends on the rate at which electrons can enter and exit the dot $G_1$ at energy $\xi_g+U_m$ and $\xi_g$ respectively. Thus, this component depends on $\gamma_g$ and can be suppressed substantially by reducing $\gamma_g$. Thus, on decreasing $\gamma_g$, the magnitude of the heat current from reservoir $G$ can be suppressed substantially. \\
\indent As demonstrated in Fig.~\ref{fig:Fig_9}(a), the triple dot setup extracts much lower heat current  from the reservoir $G$, while offering an enhanced sensitivity. In addition,  the heat current can be suppressed by a large amount without much impact on the sensitivity by decreasing $\gamma_g$. This is clearly demonstrated in Fig.~\ref{fig:Fig_9}(a) and (b), where decreasing $\gamma_g$ by a factor of $10$ in the triple dot (TD) set-up decreases the maximum extracted heat current from $1.6$fW to $0.276$fW (by a factor of almost $5.8$), while keeping the  sensitivity almost unchanged. Thus, a  smart fabrication strategy in the triple dot set-up may be employed to prevent thermometry induced refrigeration and temperature drift of the remote target reservoir $G$. 
 \section{Conclusions}\label{conclusion}
\indent To conclude, in this paper, I have proposed current based non-local thermometry as a robust and practical alternative to thermoelectric voltage based operation. Subsequently, I have investigated  current based thermometry  performance and regime of operation of the conventional dual dot set-up. Proceeding further, I have proposed a triple dot non-local thermometer which demonstrates a higher sensitivity while bypassing the need for unrealistic step-like system-to-reservoir coupling, in addition to  providing robustness against fabrication induced variability in the Coulomb coupling energy. Furthermore, it was demonstrated that suitable fabrication strategy in the triple dot set-up  aids in suppressing thermometry induced refrigeration (heat-up) and temperature drift in the remote target reservoir   to a significant extent. Thus, the triple dot set-up  hosts multitude of advantages that are necessary to deploy quantum non-local thermometers in practical applications. In this paper, I have mainly considered the limit of weak coupling which restricts electronic transport in the sequential tunneling regime and validates the use of quantum master equation for system analysis. It would, however, be interesting to investigate the impacts of cotunneling on the thermometer performance as the system is gradually tuned towards the strong coupling regime. In addition, an analysis on the impacts of electron-phonon interaction on the system performance would also constitute an interesting study. Other practical design strategies for non-local quantum thermometers  is left for future investigation. Nevertheless, the triple  dot design investigated in this paper can be employed to fabricate highly sensitive and robust non-local ``sub-Kelvin" range thermometers.  

   \indent \textbf{Acknowledgments:} Aniket Singha would like to thank financial support from Sponsored Research and Industrial Consultancy (IIT Kharagpur) via grant no. IIT/SRIC/EC/MWT/2019-20/162, Ministry of Human Resource Development (MHRD),  Government of India  via Grant No.  STARS/APR2019/PS/566/FS under STARS scheme and Science and Engineering Research Board (SERB),  Government of India  via Grant No. 
   SRG/2020/000593 under SRG scheme.\\
 % \indent \textbf{Data Availability:} The data that supports the findings of this study are available within the article
\appendix
\section{Another equivalent triple dot  set-up for efficient non-local thermometry}\label{app_a}
\begin{figure}
	\centering
	\includegraphics[width=.45\textwidth]{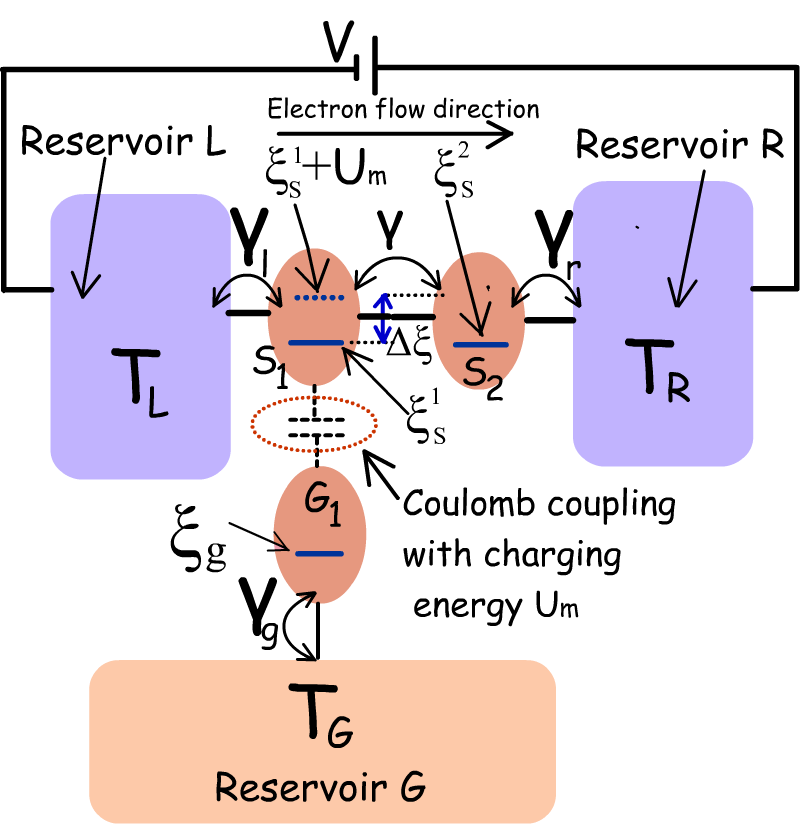}
	\caption{Schematic diagram for an equivalent triple dot  design  to accomplish efficient non-local thermometry. The thermometry performance of this set-up was found to be similar to the proposed triple dot  set-up in Fig.~\ref{fig:Fig_1}(b) with a different regime of operation. }
	\label{fig:app_1}
\end{figure}
In this section, I show a variant of the triple dot set-up that can also be employed for efficient non-local thermometry in the regime of hundreds of ``milli-Kelvin". This set-up is demonstrated in Fig.~\ref{fig:app_1} and is identical in construction to the set-up investigated in this paper (in Fig.~\ref{fig:Fig_1}.b). However, unlike the proposed set-up in Fig.~\ref{fig:Fig_1}(b), the ground-states of the two quantum dots $S_1$ and $S_2$ are aligned with each other, that is $\xi_{s}^1=\xi_{s}^2$. In this case, an electron tunneling from $G$ into $G_1$ misaligns the ground states in $S_1$ and $S_2$ and blocks the current flow through the system. A change in temperature of the reservoir $G$ impacts the probability of occupancy of $G_1$ and thus induces thermometry.  Although not elaborated here, the configuration demonstrated in Fig.~\ref{fig:app_1} demonstrates similar thermometry performance to the set-up shown in Fig.~\ref{fig:Fig_1}(b) with a different regime of operation. The configuration, demonstrated in Fig.~\ref{fig:app_1}, thus provides an alternative arrangement for efficient non-local thermometry.
\section{Derivation of quantum master equations (QME) for the triple dot thermometer}\label{app_b}
\begin{figure*}[!htb]
	\centering
	\includegraphics[width=.92\textwidth]{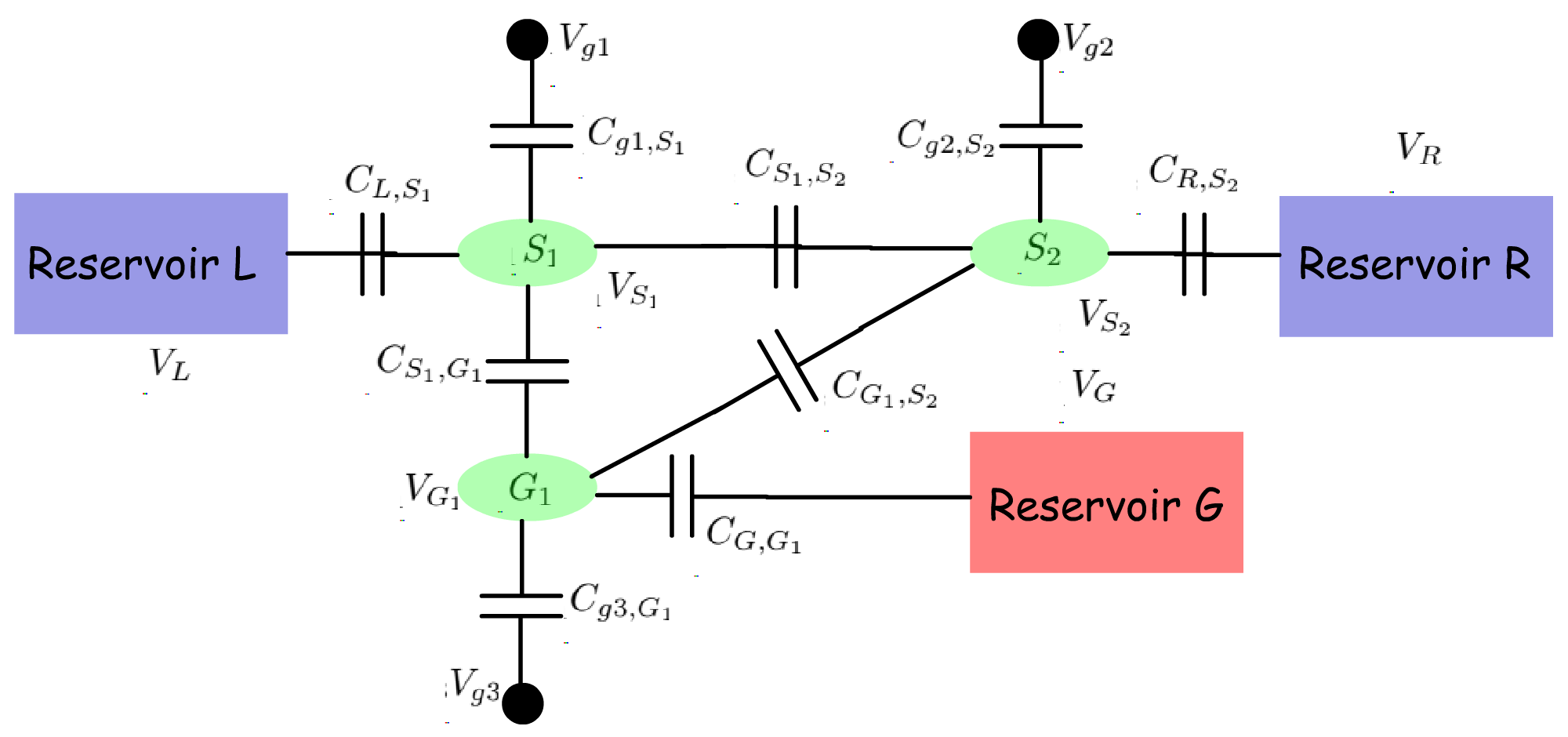}
	\caption{Schematic diagram demonstrating electrostatic interaction of the system with the adjacent electrodes and other dots. The voltages $V_{g1},~V_{g2},~V_{g3}$ are the voltages at the gate terminals of the dots $S_1$, $S_2$ and $G_1$ respectively. }
	\label{fig:app_2}
\end{figure*}
In this section,  I derive the quantum master equations (QME) for the proposed triple dot non-local thermometer, starting from the basic physics of Coulomb coupled systems. Fig.~\ref{fig:app_2} depicts  equivalent schematic model for electrostatic interaction of the quantum dots with the adjacent electrodes as well as the adjacent dots.   $V_{g1},~V_{g2},~V_{g3}$ are the voltages at the gate terminals of the dots $S_1$, $S_2$ and $G_1$ respectively. The other symbols in Fig.~\ref{fig:app_2} are self-explanatory. The potentials of the dots $S_1,~S_2$ and $G_1$ can be calculated in terms of the quantum dot charge and the  potentials at the adjacent terminals  as \cite{cb}:
\begin{widetext}
	\begin{align}
	&V_{S_1}=\frac{Q_{S_1}}{C_{S_1}^{\Sigma}}+\frac{1}{C_{S_1}^{\Sigma}}\left\{ C_{g1,S_1}V_{g1}+C_{L,S_1}V_{L}+C_{S_1,S_2}V_{S_2}+C_{S_1,G_1}V_{G_1}\right\} \nonumber \\
	&V_{G_1}=\frac{Q_{G_1}}{C_{G_1}^{\Sigma}}+\frac{1}{C_{G_1}^{\Sigma}}\left\{ C_{g3,G_1}V_{g3}+C_{G,G_1}V_{G}+C_{G_1,S_2}V_{S_2}+C_{S_1,G_1}V_{S_1}\right\} \nonumber \\
	&V_{S_2}=\frac{Q_{S_2}}{C_{S_2}^{\Sigma}}+\frac{1}{C_{S_2}^{\Sigma}}\left\{ C_{g2,S_2}V_{g2}+C_{R,S_2}V_{R}+C_{G_1,S_2}V_{G_1}+C_{S_1,S_2}V_{S_1}\right\} 
	\label{eq:poential},
	\end{align}
	where $Q_x$ is the charge in dot $x$ and the terms $C_x^{\Sigma }$ is the total capacitance seen by the dot $x$ with its adjacent environment. 
	\begin{align}
	&C_{S_1}^{\Sigma }=C_{g1,S_1}+C_{S_1,G_1}+C_{L,S_1}+C_{S_1,S_2} \nonumber \\
	&C_{S_2}^{\Sigma }=C_{g2,S_2}+C_{G_1,S_2}+C_{R,S_2}+C_{S_1,S_2} \nonumber \\
	&C_{G_1}^{\Sigma }=C_{g3,G_1}+C_{S_1,G_1}+C_{G,G_1}+C_{G_1,S_2} 
	\end{align}
	In general, each dot is  coupled strongly with its corresponding gate terminal. Thus, from a practical purposes, the effective capacitance $C_{L,S_1},~C_{G,G_1}$ and $C_{R,S_2}$ between the quantum  dots and the electrically-coupled electrodes can be neglected with respect to the gate coupling capacitances $C_{g1,S_1},~C_{g2,S_2}$ and $C_{g3,G_1}$. In addition, the dots $S_1$ and $G_1$ are strongly coupled (intentionally) by suitable fabrication techniques \cite{cap_coup_1,cap_coup_2,cap_coup3,cap_coup_4,cap_coup_5}. In addition, I assume that electrostatic coupling between $S_1-S_2$ and $G_1-S_2$ are  negligible. However,  $C_{S_1,G_1}$ is enhanced via appropriate fabrication techniques \cite{cap_coup_1,cap_coup_2,cap_coup3,cap_coup_4,cap_coup_5}, such that  $C_{S_1,G_1}>>(C_{S_1,S_2},C_{G_1,S_2})$. Hence, for  the following derivations, I  neglect the capacitances $C_{L,S_1},~C_{R,S_2},~C_{G,G_1},~C_{S_1,S_2},~C_{G_1,S_2}$. 
	Under all these  considerations, the total system electrostatic energy  can be given by \cite{cb}:
	\begin{align}
	U_{tot}=\sum_{x\in (S_1,S_2,G_1)} \frac{Q_x^2}{2C_x^{\Sigma}} +\frac{Q_{S_1}}{C^{\Sigma}_{S_1}}\left\{C_{g1,S_1}V_{g1}+C_{S_1,G_1}V_{G_1}\right\}+\frac{Q_{S_2}}{C^{\Sigma}_{S_2}}C_{g2,S_2}V_{g2}+\frac{Q_{G_1}}{C^{\Sigma}_{G_1}}\left\{C_{g3,G_1}V_{g3}+C_{S_1,G_1}V_{S_1}\right\}, \nonumber \\
	\label{eq:utot}
	\end{align}
	where it is assumed that  $C_{L,S_1},~C_{R,S_2},~C_{G,G_1},~C_{S_1,S_2},~C_{G_1,S_2}$ is negligible compared to the other capacitances in the system.
	At $0K$, the system would equilibriate at the minimum possible value of $U_{tot}$, which is termed as $U_{eq}$. The  charge $Q_{S_1}^{eq}=-qn_{S_1}^{eq},~Q_{S_2}^{eq}=-qn_{S_2}^{eq}$ and $Q_{G_1}^{eq}=-qn_{G_1}^{eq}$  in the dot $S_1,~S_2$ and $G_1$ respectively, in equilibrium (minimum energy condition)   at $0K$, can be calculated by solving the set of  equations given below:
	\begin{align}
	&\frac{\partial U_{tot}}{\partial Q_{S_1}}=\frac{Q_{S_1}}{C^{\Sigma}_{S_1}}+\frac{1}{C_{S_1}^{\Sigma}}(C_{g1,S_1}V_{g1}+C_{S_1,G_1}V_{G_1})+\frac{(C_{S_1,G_1})^2 Q_{S_1}}{C_{G_1}^{\Sigma}(C_{S_1}^{\Sigma})^2}+\frac{C_{S_1,G_1}Q_{G_1}}{C_{S_1}^{\Sigma}C_{G_1}^{\Sigma}}=0 \nonumber \\
	&\frac{\partial U_{tot}}{\partial Q_{S_2}}=\frac{Q_{S_2}}{C^{\Sigma}_{S_2}}+\frac{C_{g2,S_2}}{C_{S_2}^{\Sigma}}V_{g2} =0\nonumber \\
	&\frac{\partial U_{tot}}{\partial Q_{G_1}}=\frac{Q_{G_1}}{C^{\Sigma}_{G_1}}+\frac{1}{C_{G_1}^{\Sigma}}(C_{g3,G_1}V_{g3}+C_{S_1,G_1}V_{G_3})+\frac{(C_{S_1,G_1})^2 Q_{G_1}}{C_{S_1}^{\Sigma}(C_{G_1}^{\Sigma})^2}+\frac{C_{S_1,G_1}Q_{S_1}}{C_{G_1}^{\Sigma}C_{S_1}^{\Sigma}}=0
	\label{eq:partialderivative}
	\end{align}
	The above set of equations  can be derived by partial differentiation of Eq.~\ref{eq:utot}, and replacing  appropriate expressions obtained from partial differentiation and  algebraic manipulation of the set of Eqns.~\ref{eq:potential}.
	The number of electrons in the dots may vary stochastically  due to application of external voltage bias or thermal fluctuations from the reservoir at finite temperature. The small increase in the net  system electrostatic potential  energy due to application of external bias or thermal fluctuations from the reservoirs can be given via a Taylor's  expansion of Eq.~\eqref{eq:utot} around the equilibrium dot charges ($-qn_{S_1}^{eq},~-qn_{S_2}^{eq}$ and $-qn_{G_1}^{eq}$), along  with the condition $\frac{\partial U_{tot}}{\partial Q_{S_1}}\Big|_{Q_{S_1}=-qn_{S_1}^{eq}}=\frac{\partial U_{tot}}{\partial Q_{S_2}}\Big|_{Q_{S_2}=-qn_{S_2}^{eq}}=\frac{\partial U_{tot}}{\partial Q_{G_1}}\Big|_{Q_{G_1}=-qn_{G_1}^{eq}}=0$ as:
	\begin{eqnarray}
	U(n_{S_{1}},n_{G_{1}},n_{S_{2}})=U_{tot}-U_{eq}=\sum_{x \in (S_{1},G_{1},S_{2})}\frac{q^2}{C^{self}_{x}}\left(n_{x}^{tot}-n_x^{eq}\right)^2  +\sum_{(x_{1},x_{2})\in(S_{1},G_{1},S_{2})}^{x_1 \neq x_2} U_{x_1,x_2}\left(n_{x1}^{tot}-n_{x1}^{eq}\right)\left(n_{x2}^{tot}-n_{x2}^{eq}\right)  \nonumber \\
	\label{eq:caps}
	\end{eqnarray}
\end{widetext}
where $n_x^{tot}$ is the total number of electrons, and    $C_x^{self}$ is self capacitance of the dot $x$. $U_{x_1,x_2}$ denotes the electrostatic energy arising out of mutual  Coulomb coupling between two different quantum dots, which accounts for  a fluctuation in the electronic number of dot $x_{1(2)}$ affecting the electrostatic energy of dot $x_{2(1)}$. These quantities can be derived from the sets of Eqns.~\eqref{eq:potential}, \eqref{eq:utot} and  \eqref{eq:partialderivative}, along with the assumption $C_{L,S_1}=C_{R,S_2}=C_{G,G_1}=C_{S_1,S_2}=C_{G_1,S_2}= 0$ as:
\begin{align}
&\frac{1}{C_{S_1}^{self}}=\frac{\partial^2 U_{tot}}{\partial Q_{S_1}^2}=\frac{1}{C_{S_1}^{\Sigma}}+2\frac{(C_{S_1,G_1})^2}{(C_{S_1}^{\Sigma})^2C_{G_1}^{\Sigma}} \nonumber \\
&\frac{1}{C_{S_2}^{self}}=\frac{\partial^2 U_{tot}}{\partial Q_{S_2}^2}=\frac{1}{C_{S_2}^{\Sigma}}\nonumber \\
&\frac{1}{C_{G_1}^{self}}=\frac{\partial^2 U_{tot}}{\partial Q_{G_1}^2}=\frac{1}{C_{G_1}^{\Sigma}}+2\frac{(C_{S_1,G_1})^2}{(C_{G_1}^{\Sigma})^2C_{S_1}^{\Sigma}} \nonumber \\
&\frac{U_{S_1,G_1}}{q^2}=\frac{\partial^2 U_{tot}}{\partial Q_{G_1}\partial Q_{S_1}}=\frac{\partial^2 U_{tot}}{\partial Q_{S_1}\partial Q_{G_1}}=2\frac{C_{S_1,G_1}}{C_{S_1}^{\Sigma}C_{G_1}^{\Sigma}} \nonumber
\end{align}
\begin{align}
&\frac{U_{S_1,S_2}}{q^2}=\frac{\partial^2 U_{tot}}{\partial Q_{S_1}\partial Q_{S_2}}=\frac{\partial^2 U_{tot}}{\partial Q_{S_2}\partial Q_{S_1}}=0\nonumber \\
&\frac{U_{S_2,G_1}}{q^2}=\frac{\partial^2 U_{tot}}{\partial Q_{G_1}\partial Q_{S_2}}=\frac{\partial^2 U_{tot}}{\partial Q_{S_2}\partial Q_{G_1}}=0
\label{eq:caps_expp}
\end{align}
From Eq.~\eqref{eq:caps},  I proceed to derive the QME of the entire system. In the triple dot set-up, the additional quantum dot $S_2$ is tunnel coupled to the  $S_1$, while $G_1$ is Coulomb coupled to  $S_1$. I assume that the electrostatic energy due to   self-capacitance is much greater than the average thermal energy or the applied bias voltage, that is $E^{self}_x=\frac{q^2}{C^{self}_{x}}>> (kT,~qV)$, such that electronic transport through the Coulomb blocked energy level, due to self-capacitance, can be neglected. Thus the maximum number of electrons in the ground states of each quantum dots is limited to $1$. Under all these assumptions, the system analysis may be restricted to $2^3=8$ multi-electron  states, that I indicate by the electron number in each quantum dot. Thus a state of interest in the system may be denoted by  $\ket{n_{S_1},n_{G_1},n_{S_2}}=\ket{n_{S_1}}\tens{} \ket{n_{G_1}} \tens{} \ket{n_{S_2}}$, where $(n_{S_1},n_{G_1},n_{S_2})\in (0,1)$. To simplify these representations of  multi-electron states, with a slight abuse of notation, I rename the states as $\ket{0,0,0}\rightarrow \ket{0}$, $\ket{0,0,1}\rightarrow \ket{1}$, $\ket{0,1,0}\rightarrow \ket{2}$, $\ket{0,1,1}\rightarrow \ket{3}$, $\ket{1,0,0}\rightarrow \ket{4}$, $\ket{1,0,1}\rightarrow \ket{5}$, $\ket{1,1,0}\rightarrow \ket{6}$, and 
$\ket{1,1,1}\rightarrow \ket{7}$ \\
The simplified  Hamiltonian of the triple dot system  can hence be written as:
\begin{eqnarray}
H=\sum_{\beta}\epsilon_{\beta}\ket{\beta}\bra{\beta}+t\{\ket{3}\bra{6}+\ket{1}\bra{4}\}\nonumber \\ +U_m \{\ket{6}\bra{6}+\ket{7}\bra{7}\}+h.c.,
\end{eqnarray}
where $U_m=U^m_{S_1,G_1}$ is the electrostatic coupling energy between $S_1$ and $G_1$ in Fig.~\ref{fig:Fig_1}, $t$ denotes the interdot tunnel coupling element  or hopping parameter between $S_1$ and $S_2$ and $\epsilon_{\beta}$ is the total energy of the state $\ket{\beta}$ with respect to the vacuum state $\ket{0}$.  Under the assumption that the interdot coupling element $t$ or the reservoir to dot coupling are small, the temporal dynamics of the system density matrix can be evaluated by taking the partial trace over the entire density matrix of the combined set-up consisting of the reservoirs and the dots \cite{master_eq_1,master_eq_2,master_eq_3,master_eq_4,master_eq_5,master_eq_6}. In this framework,  the diagonal and the non-diagonal terms of the triple dot density matrix $\rho$ can be written as a set of modified Liouville equation \cite{master_eq_1,master_eq_2,master_eq_3,master_eq_4,master_eq_5,master_eq_6}:
\begin{eqnarray}
\frac{\partial \rho_{\eta \eta}}{\partial t}=-i[H,\rho]_{\eta \eta}-\sum_{\nu} \Gamma_{\eta \nu}\rho_{\eta \eta}+\sum_{\delta}\Gamma_{\delta \eta }\rho_{\delta \delta} \nonumber \\
\frac{\partial \rho_{\eta \beta}}{\partial t}=-i[H,\rho]_{\eta \beta}-\frac{1}{2}\sum_{\nu} \Big(\Gamma_{\eta \nu}+\Gamma_{\beta \nu }\Big) \rho_{\eta \beta}, \nonumber \\
\label{eq:time_derivate}
\end{eqnarray}
where  $\rho_{\eta \beta}=\bra{\eta}\rho \ket{\beta}$ and $[x,y]$ denotes the commutator of the operators $x$ and $y$. The  terms $\rho_{\eta \eta}$ and $\rho_{\eta \beta}$ in the above equation represent  diagonal and non-diagonal elements of the system density matrix respectively. The off-diagonal elements $\rho_{\eta \beta}$ account for coherent inter-dot tunneling, in addition to tunneling of electrons between the dots and the reservoirs. The off-diagonal terms $\rho_{\eta \beta}$, thus, are only non-zero and finite when electron tunneling can result in the transition between the states $\eta$ and $\beta$ or vice-versa. The parameters $\Gamma_{xy}$ account for the transition between system   states  due to electronic tunneling between the system and the reservoirs and are only finite when the system state transition from  $\ket{x}$ to $\ket{y}$ (or vice-versa)  is possible due to electron transfer between the system and the reservoirs. Assuming  a statistical quasi-Fermi distribution  inside the reservoirs,  $\Gamma_{xy}$  can be given as:   
\begin{eqnarray}
\Gamma_{xy }=\gamma_{\lambda}f_{\lambda}(\epsilon_{y}-\epsilon_{x}), %\nonumber \\
% \Gamma_{\delta \eta  }=\frac{\gamma_{\delta \eta }}{h}f({\epsilon_{\eta}}-\epsilon_{\delta}) \nonumber \\
\label{eq:c4}
\end{eqnarray}
where   $f_{\lambda}(\epsilon)$ denotes the probability of occupancy of an electron in the corresponding reservoir $\lambda$ (driving the state transition) at energy $\epsilon$, $\epsilon_{x (y)}$ is the total electronic energy  in the state $\ket{x(y)}$ compared to  vacuum, and $\gamma_{\lambda}$ denotes the system to reservoir coupling for the corresponding reservoir $\lambda$.  \\
For the triple dot set-up,  tunneling of electrons between the quantum dots drives the system  from $\ket{4}$ to $\ket{1}$  and from $\ket{3}$ to $\ket{6}$ (or vice-versa). In steady state, the time-derivative of each   density matrix element $[\rho]$ vanishes. Hence, employing the second equation of \eqref{eq:time_derivate}, I get, 
\begin{equation}
\rho_{4,1}=\rho^{*}_{1,4}=\frac{\rho_{4,4}-\rho_{1,1}}{\epsilon_{4}-\epsilon_{1}-i\frac{\Upsilon_{4,1}}{2}}
\label{eq:c5}
\end{equation}
\begin{equation}
\rho_{6,3}=\rho^{*}_{3,6}=\frac{\rho_{6,6}-\rho_{3,3}}{\epsilon_{6}-\epsilon_{3}-i\frac{\Upsilon_{6,3}}{2}},
\label{eq:c6}
\end{equation}
where $\Upsilon_{x,y}$ is the sum of net tunneling rates between the system and the reservoirs that leads to the decay of the states $\ket{x}$ and $\ket{y}$. In Eqns.~\eqref{eq:c5} and \eqref{eq:c6}, $\Upsilon_{4,1}$ and $\Upsilon_{6,3}$ are given by:
\begin{gather}
\Upsilon_{4,1}=\Gamma_{\ket{4},\ket{0}}+\Gamma_{\ket{4},\ket{6}}+\Gamma_{\ket{4},\ket{5}}+\Gamma_{\ket{1},\ket{0}}+\Gamma_{\ket{1},\ket{6}}+\Gamma_{\ket{1},\ket{3}} \nonumber \\
\Upsilon_{6,3}=\Gamma_{\ket{6},\ket{4}}+\Gamma_{\ket{6},\ket{2}}+\Gamma_{\ket{6},\ket{7}}+\Gamma_{\ket{3},\ket{1}}+\Gamma_{\ket{3},\ket{2}}+\Gamma_{\ket{3},\ket{7}} 
\end{gather}
From Eq.~\eqref{eq:time_derivate}, the time derivative of the density matrix elements $\rho_{6,6}$ and $\rho_{3,3}$ can be given by:
\begin{align}
\dot{\rho}_{6,6}=&it(\rho_{6,3}-\rho_{3,6})-\left(\Gamma_{\ket{6},\ket{4}}-\Gamma_{\ket{6},\ket{2}}+\Gamma_{\ket{6},\ket{7}}\right)\rho_{6,6}\nonumber \\&+\Gamma_{\ket{4},\ket{6}}\rho_{4,4}+\Gamma_{\ket{2},\ket{6}}\rho_{2,2}+\Gamma_{\ket{7},\ket{6}}\rho_{7,7} \nonumber \\
\dot{\rho}_{4,4}=&it(\rho_{4,1}-\rho_{1,4})-\left(\Gamma_{\ket{4},\ket{0}}-\Gamma_{\ket{4},\ket{6}}+\Gamma_{\ket{4},\ket{5}}\right)\rho_{4,4}\nonumber \\&+\Gamma_{\ket{0},\ket{4}}\rho_{0,0}+\Gamma_{\ket{6},\ket{4}}\rho_{6,6}+\Gamma_{\ket{5},\ket{4}}\rho_{5,5} \nonumber \\
\label{eq:b8}
\end{align}
Substituting the values of  $\rho_{6,3},~\rho_{3,6},~\rho_{4,1}$ and $\rho_{1,4}$  from Eq.~\eqref{eq:c5} and \eqref{eq:c6} in Eq.~\eqref{eq:b8}, time derivative of the probability of the states $\ket{4}$ and $\ket{6}$ can be given by:
\begin{eqnarray}
\dot{p_6}=\dot{\rho}_{6,6}=\sum_{\alpha=}\left(  -\Gamma_{\ket{6},\ket{\alpha}}p_6+\Gamma_{\ket{\alpha},\ket{6}}p_{\alpha} \right) \nonumber \\
-\Lambda_{\ket{6},\ket{3}}p_6+\Lambda_{\ket{3},\ket{6}}p_3
\end{eqnarray}\\
\begin{eqnarray}
\dot{p_4}=\dot{\rho}_{4,4}=\sum_{\alpha=}\left(  -\Gamma_{\ket{4},\ket{\alpha}}p_3+\Gamma_{\ket{\alpha},\ket{4}}p_{\alpha} \right) \nonumber \\
-\Lambda_{\ket{4},\ket{1}}p_4+\Lambda_{\ket{1},\ket{4}}p_1,
\end{eqnarray}
where $p_{\eta}=\rho_{\eta,\eta}$ and 
\begin{gather}
\Lambda_{\ket{6},\ket{3}}=\Lambda_{\ket{3},\ket{6}}=t^2\frac{\Upsilon_{6,3}}{(\epsilon _6-\epsilon _3)^2+\frac{\Upsilon _{6,3}^2}{4}}\nonumber \\
\Lambda_{\ket{4},\ket{1}}=\Lambda_{\ket{1},\ket{4}}=t^2\frac{\Upsilon_{4,1}}{(\epsilon _4-\epsilon _1)^2+\frac{\Upsilon _{4,1}^2}{4}} \nonumber \\
\label{eq:tun_rate}
\end{gather}
In Eq.~\ref{eq:tun_rate}, $\Lambda_{\ket{4},\ket{1}}$ and $\Lambda_{\ket{6},\ket{3}}$ denote  the rates of interdot tunneling  between $S_1$ and $S_2$ with empty and occupied ground states of  $G_1$  respectively. By a smart choice of the ground state energy positions,  the condition $\epsilon_6=\xi_g+\xi_s^1+U_m=\xi_g+\xi_s^2=\epsilon_3$, that is  $\xi_s^2=\xi_s^1+U_m$ is satisfied. In such a case, under the condition $U_m>>|\Upsilon _{4,1}|$, I get $\Lambda_{\ket{6},\ket{3}}>>\Lambda_{\ket{4},\ket{1}}$. This  condition implies that the inter-dot tunneling probability between $S_1$ and $S_2$  is negligible when the ground state  in $G_1$ is empty.\\
\indent For the calculation of current, we need to know the probability of ground state in $S_1$ or $S_2$. Since, the electronic transport via the ground states in $S_1$ and $G_1$ are coupled to each other  by Coulomb interaction, I consider $S_1$ and $G_1$ as a  sub-system ($\varsigma_1$) of the total triple dot system. $S_2$ is considered to be the complementary sub-system ($\varsigma_2$) of the entire set-up. For the range of parameters used in this case, the condition $U_m>>|\Upsilon _{4,1}|$ is satisfied, which leads to $\Lambda_{\ket{4},\ket{1}}<<\Lambda_{\ket{6},\ket{3}}$. Hence, to simplify the calculations, I assume that   $\Lambda_{\ket{4},\ket{1}}\approx 0$ all practical purposes relating to electron transport. In the following discussion, I simply denote $\Lambda_{\ket{6},\ket{3}}$ as $\gamma$ to represent the interdot tunnel coupling. I denote the  occupancy probability  of the subsystem $\varsigma_1$ as $P_{i,j}^{\varsigma_1}$, where $i$ and $j$ denote the electron number in the ground state of $S_1$ and $G_1$ respectively, while $P_k^{\varsigma_2}$  denotes the ground state occupancy probability of $S_2$. Note that splitting the entire system into two sub-systems in this fashion demands the limit of weak tunnel and Coulomb coupling between the two sub-systems such that the state of one sub-system remains unaffected by the state of the complementary sub-system. In such a limit, we can write $\rho_{0,0}=P^{\varsigma_1}_{0,0}P^{\varsigma_2}_{0},~\rho_{1,1}=P^{\varsigma_1}_{0,0}P^{\varsigma_2}_{1},~\rho_{2,2}=P^{\varsigma_1}_{0,1}P^{\varsigma_2}_{0},~\rho_{3,3}=P^{\varsigma_1}_{0,1}P^{\varsigma_2}_{1},~\rho_{4,4}=P^{\varsigma_1}_{1,0}P^{\varsigma_2}_{0},~\rho_{5,5}=P^{\varsigma_1}_{1,0}P^{\varsigma_2}_{1},~\rho_{6,6}=P^{\varsigma_1}_{1,1}P^{\varsigma_2}_{1},~\rho_{7,7}=P^{\varsigma_1}_{1,1}P^{\varsigma_2}_{1}$ The quantum master equations (QME) for the sub-system $\varsigma_1$ can be given in terms of two or more diagonal elements of the density matrix, in \eqref{eq:time_derivate}, as:
\begin{widetext}
	\begin{align}
	\frac{d}{dt}(P_{0,0}^{\varsigma_1})=\frac{d}{dt}\left( \rho_{0,0}+\rho_{1,1}\right)=&- P_{0,0}^{\varsigma_1}\{\gamma_l f_L(\xi_s^1)+\gamma_g f_G(\xi_g)\}+\gamma_g P_{0,1}^{\varsigma_1}\{1-f_G(\xi_g)\}+\gamma_l P_{1,0}^{\varsigma_1}\{1-f_L(\xi_s^1)\}\nonumber \\
	\frac{d}{dt}(P_{1,0}^{\varsigma_1})=\frac{d}{dt}\left(\rho_{5,5}+\rho_{4,4}\right)=&- P_{1,0}^{\varsigma_1}\left\{\gamma_l\left(1-f_L(\xi_{s}^1)\right)+\gamma_g f_G(\xi_g+U_m)\right\}+\gamma_g P_{1,1}^{\varsigma_1}\left\{1-f_G(\xi_g+U_m)\right\}+\gamma_g P_{0,0}^{\varsigma_1}f_L(\xi_{s}^1) \nonumber \\
	\frac{d}{dt}(P_{0,1}^{\varsigma_1})=\frac{d}{dt}\left(\rho_{2,2}+\rho_{3,3}\right)=&-P_{0,1}^{\varsigma_1}\left\{\gamma_g \left(1-f_g(\xi_{g}^1)\right)+\gamma_l f_L(\xi_s^1+U_m)+{\gamma} P^{\varsigma_2}_1\right\} \nonumber \\& +\gamma_g P_{0,0}^{\varsigma_1}f_G(\xi_g)+P_{1,1}^{\varsigma_1}\left\{\gamma_l \left(1-f_L(\xi_s^1+U_m)\right)+\gamma P^{\varsigma_2}_{0}\right\} \nonumber \\
	\frac{d}{dt}(P_{1,1}^{\varsigma_1})=\frac{d}{dt}\left(\rho_{7,7}+\rho_{6,6}\right)=&-P_{1,1}^{\varsigma_1}\left\{\gamma_g\left(1-f_g(\xi_{g}^1+U_m)\right)+\gamma_l \left(1-f_L(\xi_s^1+U_m)\right)+\gamma P^{\varsigma_2}_0\right\} \nonumber \\ &+\gamma_g P_{1,0}^{\varsigma_1}f_G(\xi_g+U_m) +P_{0,1}^{\varsigma_1}\left\{\gamma_l f_L(\xi_s^1+U_m)+\gamma P^{\varsigma_2}_{1}\right\} \nonumber \\
	\label{eq:first_sys1}
	\end{align} 
	where $\gamma=\Lambda_{\ket{6},\ket{3}}=\Lambda_{\ket{3},\ket{6}}$ and $\Lambda_{\ket{4},\ket{1}}=\Lambda_{\ket{1},\ket{4}}=0$.  I  assume quasi Fermi-Dirac electron distribution at the reservoirs. Hence,  corresponding to the reservoir $\lambda$, and $\lambda \in (L,R,G)$ $f_{\lambda}(\epsilon)=\left\{1+exp\left(\frac{\epsilon-\mu_{\lambda}}{kT_{\lambda}}\right)\right\}^{-1}$.
	Similarly, the QME of the sub-system $\varsigma_2$ can be written as:
	\begin{align}
	&\frac{d}{dt}(P_{0}^{\varsigma_2})=\frac{d}{dt}\left( \rho_{6,6}+\rho_{4,4}+\rho_{2,2}+\rho_{0,0}\right)=-P_{0}^{\varsigma_2}\{\gamma_r f_R(\xi_s^2)+{\gamma}P_{1,1}^{\varsigma_1}\}+P_1^{\varsigma_2}\{\gamma_r \left(1-f_R(\xi_{s}^2)\right)+{\gamma}P^{\varsigma_1}_{0,1}\}\nonumber \\
	&\frac{d}{dt}(P_{1}^{\varsigma_2})=\frac{d}{dt}\left( \rho_{7,7}+\rho_{5,5}+\rho_{3,3}+\rho_{1,1}\right)=-P_1^{\varsigma_2}\{\gamma_r  \left(1-f_R(\xi_{s}^2)\right)+{\gamma}P^{\varsigma_1}_{0,1}\}+P_{0}^{\varsigma_2}\{\gamma_r f_R(\xi_s^2)+{\gamma}P_{1,1}^{\varsigma_1}\}\nonumber \\
	\label{eq:second_sys1}
	\end{align}
	 The L.H.S of Eqns.~\eqref{eq:first_sys1} and \eqref{eq:second_sys1} are zero in steady state. 	 The set of Eqns.~\eqref{eq:first_sys1} and \eqref{eq:second_sys1} form a coupled system of equations which were solved  iteratively via Newton-Raphson method. On solution of the steady-state probabilities, the charge current $I_{L(R)}$ between reservoir $L$ and $R$  and the electronic heat current ($I_{Qe}$) extracted from the reservoir $G$ can be calculated by the equations:
	\begin{equation}
	I_L= q \gamma_l \times \left\{P^{\varsigma_1}_{0,0}f_L(\xi_s^1)+P^{\varsigma_1}_{0,1}f_L(\xi_s^1+U_m)-P^{\varsigma_1}_{1,0}\{1-f_L(\xi_s^1)\}-P^{\varsigma_s^1}_{1,1}\{1-f_L(\xi_s^1+U_m)\}\right\} 
	\end{equation}
	\begin{equation}
	I_R= -q\gamma_r \times \left\{P^{\varsigma_2}_{0}f_R(\xi_s^1)-P^{\varsigma_2}_{1}\{1-f_R(\xi_s^1)\}\right\} 
	\end{equation}
	\begin{align}
	I_{Q}=
	\gamma_g \times \left\{(\xi_g+U_m-\mu_g)\left\{P^{\varsigma_1}_{1,0}f_G(\xi_g+U_m)-P^{\varsigma_1}_{1,1}\{1-f_G(\xi_g+U_m)\}\right\} \right\}\nonumber \\
	+ \gamma_g \times\left\{(\xi_g-\mu_g)\times \left\{P^{\varsigma_1}_{0,0}f_G(\xi_g)  -P^{\varsigma_n}_{0,1}\{1-f_G(\xi_g)\}\right\}\right\}  
	\label{eq:heat2}
	\end{align}
	Since,  net current into (out-of) the reservoir $G$ is zero, we have 
	\begin{equation}
	I_G=
	q\gamma_g \times \left\{P^{\varsigma_1}_{1,0}f_G(\xi_g+U_m)-P^{\varsigma_1}_{1,1}\{1-f_G(\xi_g+U_m)\}+P^{\varsigma_1}_{0,0}f_G(\xi_g)  -P^{\varsigma_n}_{0,1}\{1-f_G(\xi_g)\}\right\}=0
	\label{eq:gate_curr}
	\end{equation}
	Substituting Eq.~\eqref{eq:gate_curr} in Eq.~\eqref{eq:heat2}, I get
	\begin{equation}
	I_{Q}=\gamma_g \times U_m\left\{P^{\varsigma_1}_{1,0}f_G(\xi_g+U_m)-P^{\varsigma_1}_{1,1}\{1-f_G(\xi_g+U_m)\}\right\} 
	\end{equation}
\end{widetext}

 \bibliography{apssamp}
\end{document}